\documentclass[a4paper,11pt]{article}
\pdfoutput=1 

\usepackage{jcappub} 

\usepackage[T1]{fontenc} 
\usepackage{epsfig}
\usepackage{epstopdf,units, color}

\title{Anisotropy test of the axion-like particle Universe opacity effect: \\ 
a case for the Cherenkov Telescope Array}

\author[a]{Denis Wouters,}
\author[a]{Pierre Brun,}

\affiliation[a]{CEA Irfu, Centre de Saclay, F-91191 Gif-sur-Yvette France}

\emailAdd{denis.wouters@cea.fr}
\emailAdd{pierre.brun@cea.fr}

\abstract{

The Universe opacity to gamma rays is still an open question, in particular anomalies may have been observed. Assuming that such anomalies find their origin in conventional physics like intrinsic source spectra or the density of the extragalactic background light, they would be evenly distributed over the sky. If they exist, axion-like particles (ALPs) would have a potential effect on the opacity of the Universe to gamma rays, possibly related to the anomalies in the spectral indices of distant gamma-ray sources. In the scenario where ALPs from distant sources convert back to photons in the Galactic magnetic field, their effect on the opacity is expected to depend on the position of the sources. In that case the anomaly is expected to exhibit peculiar correlations on the sky. We propose a method to test the origin of the opacity anomaly, based on angular correlations of spectral softening anomalies.  Such a diagnosis requires a wide-field survey of high-energy gamma-ray sources over a broad range of energy. The future Cherenkov Telescope Array (CTA) is perfectly suited to perform such a study. It is shown that while the current sample of sources is not large enough to base conclusions on, with this method CTA will be sensitive to ALP couplings to gamma rays of the order of $3\times 10^{-11} \;\rm GeV^{-1}$ for ALP masses below $10^{-8}$ eV.
}

\begin{document}
\maketitle
\flushbottom

\section{Introduction}
\label{sec:intro}

Axion-like particles (ALPs) are hypothetical particles that appear in extensions of the standard model of particle physics~\cite{Peccei:1977hh,1987PhR...150....1K,2012PDU.....1..116R}. They are light pseudo-scalar bosons with a two-photon coupling. This coupling leads to a mixing of photon and ALP states when propagation in an external magnetic field is considered~\cite{1988PhRvD..37.1237R}. This mixing is used to perform searches for ALPs in various ways, considering ALPs thermally produced in the Sun~\cite{2011PhRvL.107z1302A}, ALPs produced with Lasers~\cite{Ehret:2010mh}, or ALP dark matter~\cite{Asztalos:2009yp}. In this article we focus on a possible astrophysical signature of ALPs in the field of $\gamma$-ray astronomy. Astrophysical environments do offer magnetic fields over large volumes, which can be in some cases ideal conditions for observing the $\gamma$-ray/ALP mixing. In particular, the existence of ALPs could modify the opacity of the Universe to $\gamma$ rays~\cite{2003JCAP...05..005C,2007PhRvD..76l1301D,2008PhRvD..77f3001S,Mirizzi:2009aj,2009PhRvD..79l3511S,DeAngelis:2011id,2013PhRvD..87c5027M}. This fact raised interest in the last decade after the observation of distant high-energy photon sources having hard spectral indices~\cite{2000PhLB..493....1P,2006Natur.440.1018A}. $\gamma$ rays are expected to interact with the extragalactic background light (EBL) which fills the Universe in the optical and infrared bands~\cite{1967PhRv..155.1408G,2013APh....43..112D}. Above some threshold, they are absorbed as they produce $e^\pm$ pairs. In the Earth frame, this threshold energy for $\gamma$ rays falls between 100 GeV and a few TeV for sources at redshifts between 0.1 and 0.5. This energy-dependent effect implies a softening of the spectral indices of such distant sources. The observation of hard spectrum sources and distant sources by ground-based imaging atmospheric Cherenkov telescopes (IACTs) such as HEGRA~\cite{1996A&A...311L..13P}, HESS~\cite{2006A&A...457..899A}, MAGIC~\cite{2006ApJ...637L..41A} and VERITAS~\cite{2002APh....17..221W} is considered in some studies as a possible hint for new physics. One conventional way to solve this puzzle is to revise the EBL models, mainly by lowering the photon density down to values that are close to lower limits set by galaxy counts~\cite{2010A&A...515A..19K}. Recently, the observation of the spectral features induced by pair production led to measurements of the EBL level at different redshifts by HESS~\cite{2013A&A...550A...4H} and Fermi~\cite{2012Sci...338.1190A}, showing a possible consistent picture. The solution to the opacity puzzle is still controversial though, as for instance the authors of~\cite{Horns:2012fx} claim for the persistence of the anomaly even with new EBL models. One possible explanation of the opacity anomaly is provided by the ALP hypothesis~\cite{2007PhRvD..76l1301D,2013PhRvD..87c5027M}. In that case the $\gamma$ rays traveling disguised as ALPs for a fraction of their path, they are less efficient to produce $e^\pm$ pairs. If the ALPs convert back to photons before they reach the Earth, the high-energy $\gamma$ rays in excess can give rise to the anomaly. The mixing between photons and ALPs can occur in any magnetic field that exist on the line of sight: around the source itself, in the surrounding structure (e.g. the galaxy cluster), in the intergalactic medium, or in the Milky Way. Still for their tendency to reduce the opacity to $\gamma$ rays, ALPs have been suggested to be at the origin of a possible mechanism to explain the very high-energy emission of some class of blazars (namely the Flat-Spectrum Radio Quasars, FSRQs)~\cite{2012PhRvD..86h5036T,2013arXiv1306.5865M}, although some conventional scenarios might explain it too~\cite{2013arXiv1307.1779B}. For both EBL absorption and absorption in FSRQs, the ALP parameters are invoked in the same region of the parameter space. Concerning the extragalactic opacity, as argued in~\cite{2008PhRvD..77f3001S,Horns:2012kw} the most natural scenario might be that the $\gamma$ rays convert into ALPs close to the source, and convert back to $\gamma$ rays in the Milky-Way magnetic field. This is the scenario we envisage in this paper. Here, we note that if this specific scenario was to explain some opacity anomaly, then the magnitude of the anomaly would depend on the value of the galactic magnetic field (GMF) along specific lines of sight. Conversely, if the anomaly is related to the sources themselves, or to the EBL, there is no reason why it would correlate with the GMF, it would rather be evenly distributed on the sky. In the following, we use the spectral break between the high-energy and the very high-energy bands to quantify the anomaly. Although the sky pattern of the GMF integrated values may depend on the considered GMF model, it turns out that the angular dependence of the anomaly autocorrelation on the sky shows similar trends when ALPs are involved, whatever the model. In every GMF model, if ALP-induced, the anomaly is correlated on small scales on the sky and anti-correlated on large scales. This is the test we propose to perform with the Cherenkov Telescope Array (CTA), which is the only foreseen instrument capable of detecting a sufficient number of sources at both HE and VHE for this test to be conclusive. Indeed as we shall see, the current sample of VHE extragalactic sources is too sparse to be constraining. If positive in the future, {\it i.e.} if CTA shows that the anomaly is excessively correlated on small scales and anti-correlated on large scales, such a signal could be interpreted as a strong indication for the existence of ALPs.

The paper is organized as follows. First we recall the basic description of the absorption mechanism on the EBL and the way ALPs can lead to an excess of transparency. In Sec.~\ref{sec:2}, details are given concerning the scenario we propose to test, in terms of ALP parameters and physical regions where the mixing occurs. Then in Sec.~\ref{sec:3}, the formal derivation of the test variable is presented together with simulation results showing the expected ALP signal. Sec.~\ref{sec:4} presents the results of the proposed test applied to currently available data. In Sec.~\ref{sec:5}, the method for producing expected samples of extragalactic sources with CTA is described. Then the CTA sensitivity to ALPs using angular correlations is derived in Sec.~\ref{sec:6}. The results are eventually discussed in Sec.~\ref{sec:7}.

\section{Universe opacity to $\gamma$ rays and ALPs}
\label{sec:1}

\subsection{Conventional derivation of the attenuation}

Gamma rays can interact with cosmic photon backgrounds and produce pairs of electrons and positrons of mass $m_{\rm e}$. At 1 TeV, the pair production threshold is $m_{\rm e}^2/\text{1 TeV}\sim 0.26\rm \; eV$, meaning that $\gamma$ rays are absorbed by pair creation in collisions with photons in the infrared band~\cite{1967PhRv..155.1408G}. This light is the EBL, it originates from stars and interstellar dust emission, and fills the Universe~\cite{2001ARA&A..39..249H}. Because of this pair-creation process, the optical depth $\tau$ for $\gamma$ rays is not zero and their flux is attenuated according to 
\begin{equation}
\phi_{\rm obs}(E_\gamma)\;\;=\;\;\phi_{\rm s} ( E_\gamma)\;\times\; \exp\left (-\tau(E_\gamma, z_{\rm s}) \right )\;\;,
\label{eq:attenuation}
\end{equation}
where $\phi_{\rm s} $ is the intrinsic source flux at redshift $z_{\rm s}$ and $\phi_{\rm obs}$ is the observed flux. The optical depth depends on the cosmological parameters through the distance-redshift relation $\ell(z)$, the EBL density at redshift $z$ and energy $\epsilon$ $n_{\rm EBL}(z,\epsilon)$ and the angle-averaged pair production cross section $\sigma_{\gamma\gamma}$ as
\begin{equation}
\tau(E_\gamma, z_{\rm s})\;\;=\;\;\int_0^{z_{\rm s}} d \ell(z) \int_{\epsilon_{\rm th}}^\infty d\epsilon\; \sigma_{\gamma\gamma} (E_\gamma, \epsilon)\;n_{\rm EBL}(z,\epsilon)\;\;.
\label{eq:ebl}
\end{equation}
The cosmological parameters used in $\ell(z)$ are taken from the latest results from {\it Planck} for a spatially-flat $\Lambda$CDM universe with $H_0=67.3\;\rm km/s/Mpc$, $\Omega_m=0.315$ and $\Omega_\Lambda=0.685$~\cite{2013arXiv1303.5076P}. The pair-production process results in a cut off in source spectra, similar to an exponential cut off with an additional wiggle. This wiggle is due to the double-bump structure of the EBL spectral energy density in the optical~\cite{1999A&A...349...11A,2003A&A...403..523A}. The position of the cut off in the energy spectrum depends on the redshift of the source. An example of such an absorption feature is shown later in Fig.~\ref{fig:absorption} in the case of a source at a redshift of $z=0.1$ and an EBL model from~\cite{2008A&A...487..837F}. The EBL density and its variation with redshift is very difficult to measure directly because of foreground contamination from the Milky Way, and is subject to large uncertainties. Estimations based on integrated galaxy counts~\cite{2000MNRAS.312L...9M,2004ApJS..154...39F,2006A&A...451..417D} give lower limits on the density of the EBL. Recently the H.E.S.S. collaboration used $\gamma$-ray absorption to perform a measurement of the EBL density~\cite{2013A&A...550A...4H}. A similar study has been performed by the {\it Fermi}-LAT collaboration at higher redshift~\cite{2012Sci...338.1190A}. Without these measurements, or if one wishes to infer more precise effects, it is necessary to rely on models for the evolution of the EBL density with redshift (for a review, see~\cite{2013APh....43..112D}). 

As previously mentioned, the observation of sources at redshifts up to $\sim$0.5 with IACTs challenged this interpretation of the extragalactic absorption~\cite{2000PhLB..493....1P,2006Natur.440.1018A,2008Sci...320.1752M}. Although the recent measurements by H.E.S.S. (see also~\cite{2013A&A...554A..75S}) seem not in conflict with the most recent EBL models, some studies still show evidence for a pair-production anomaly~\cite{Horns:2012fx,2013PhRvD..87c5027M}. In~\cite{Horns:2012fx} for instance, the authors claim that data points corresponding to large optical depths are in tension with EBL models. It corresponds to the fact that the VHE part of the extragalactic source spectra are too hard compared to the HE parts. A diagnostic tool for this effect is the difference of spectral indices (absolute values) between HE and VHE $\Delta \Gamma = \Gamma_{\rm VHE} -  \Gamma_{\rm HE}$. $\Delta\Gamma$ is therefore a positive number that is too small in case of an anomaly. The conventions regarding the energy bands and the use of this parameter are given in Sec.~\ref{sec:3}. Because the pair-production process depends on the energy, it could be envisaged that this effect is due to a lower EBL density, or --as $\tau\propto {\rm c}/H_0$-- to a Hubble constant that is greater than usually assumed~\cite{2011A&A...536A..18P, 2013ApJ...771L..34D}. In those two cases, the anomaly is expected to be the same wherever the source is located on the sky. Indeed the Hubble rate is isotropic at the few percent level~\cite{2013PhRvD..87l3522C}. Observed from the Earth, the EBL is isotropic at the 15\% level on angular scales smaller than $2^\circ$~\cite{2011A&A...536A..18P,2013ApJ...772...77V}. A dipole of the order of 1\% is expected~\cite{2005PhR...409..361K}. Anyway, an anisotropy of the transparency anomaly would be caused by a variation of the integrated value of the EBL density over the whole line of sight. This is quite unlikely to be sizable as the EBL is expected to be homogeneous over the cosmological distances under consideration here. In any case then, the variation of the anomaly on the sky should not be larger than what is plausible from intrinsic fluctuations from source to source. Another interesting possibility is that ALPs mix with photons in such a way that the Universe happens to be more transparent. As we shall see, in that case specific anomaly patterns can be expected.

\subsection{How ALPs modify the transparency to $\gamma$ ray}

ALPs are expected to interact with photons through the term
\begin{equation}
\mathcal{L}_{\gamma a}\;\;=\;\;-\frac{1}{4}\;g_{\gamma a}\;F_{\mu\nu}\tilde{F}^{\mu\nu} \;a \;\;=\;\; g_{\gamma a}\;\vec{E}\cdot\vec{B}\; a\;\;,
\end{equation}
where $g_{\gamma a}$ is the ALP-photon coupling strength (in $\rm GeV^{-1}$), $F$ is the electromagnetic tensor, $\tilde{F}$ its dual, and $a$ is the ALP field. It can be shown that in the presence of an external magnetic field, the ALP state mix with the two polarization states of the photon, such that the propagation eigenstates are a mix of ALP and photon~\cite{1988PhRvD..37.1237R}. In that case, the propagation of the photon/ALP system can be described by the following equation
\begin{equation}
\left(E - i\partial_z + \mathcal{M} \right )
\left(\begin{array}{c} A_1 \\ A_2 \\ a \end{array} \right) = 0\;\;,
\label{eq:propag}
\end{equation}
where $A_1$ and $A_2$ are the photon polarization amplitudes, and $\mathcal{M}$ is the mixing matrix:
\begin{equation}
\mathcal{M}\;\;=\;\;\left(\begin{array}{ccc} \Delta_{11}-i\Delta_{\mathrm{abs}} & 0 & \Delta_\mathrm{B}\cos\phi \\ 0 & \Delta_{11}-i\Delta_{\mathrm{abs}} & \Delta_\mathrm{B}\sin\phi \\ \Delta_\mathrm{B}\cos\phi & \Delta_\mathrm{B}\sin\phi\ & \Delta_\mathrm{a} \end{array}\right)\;\;,
\label{eq:matrix}
\end{equation}
where:

\begin{eqnarray}
&\Delta_{11}&=\; -\frac{m_\gamma^2}{2 E}\;\;\text{accounts for the effective mass of the photon, }\\
&\Delta_{a}&=\; -\frac{m_{\rm a}^2}{2 E}\;\;\text{accounts for the ALP mass,}\\
&\Delta_{B}&=\; \frac{g_{\gamma a}\;B_t}{2}\;\;\text{is the $\gamma$/ALP mixing term, and}\\
&\Delta_{\rm abs}&=\; \frac{\tau}{ 2s}\;\;\text{accounts for the EBL absorption}.\label{eq:abs}
\end{eqnarray}

The photon can acquire an effective mass in astrophysical plasmas, such that $m^2_\gamma=4\pi \alpha \,n_{\rm e}/ m_{\rm e}$, with $\alpha$ being the fine structure constant and $n_{\rm e}$ the electron density. $m_{\rm a}$ is the ALP mass and $B_t$ is the value of the strength of the magnetic field projected on the transverse-to-propagation plane. $\phi$ is the angle defined by the direction of $\vec{B_t}$ and the first photon polarization component. The $i\Delta_{\rm abs}$ term is imaginary, so that the matrix is not Hermitian and the propagation process does not preserve unitarity. The sign of this term in $\mathcal{M}$ depends on the sign convention used for the derivative in Eq.~\ref{eq:propag}. In the expression~\ref{eq:abs}, $\tau$ is the optical depth corresponding to Eq.~\ref{eq:attenuation} and $s$ is the size of the physical region over which the attenuation is computed.

To derive the relevant orders of magnitudes, one can consider a single magnetic domain. In that case, the problem is reduced to a 2-dimensional mixing with a mixing angle $\theta$ such that
\begin{equation}
\tan 2\theta\;=\;\frac{2\,g_{\gamma a}\,B_t\,E}{m_{\rm a}^2}\;\;.
\end{equation}
In that case the probability for a photon polarized along the transverse component of the magnetic field to convert into ALP after a distance $\ell$ is
\begin{equation}
P_{\gamma\rightarrow a}\;\;=\;\;\sin^22\theta \sin^2\left ( \frac{2\pi \ell}{\lambda(E) }\right )\;\;.
\label{eq:proba1}
\end{equation}
If $\theta$ is small, the mixing is not efficient and photons propagate normally. If $\theta$ is close to $\pi/4$ though, the mixing is strong and the propagation of photons is affected. Defining the critical energy above which the mixing is strong as~\cite{2008PhLB..659..847D}
\begin{equation}
E_{\rm c}\;\;=\;\;\frac{|m_{\rm a}^2 - m_{\gamma}^2|}{2\, g_{\gamma a} \,B_t}\;\;,
\label{eq:Ec}
\end{equation}
the oscillation length can be written
\begin{equation}
\lambda(E)\;\;=\;\;\frac{4 \pi}{g_{\gamma a}\,B_t\,\sqrt{1+\left ( \frac{E_{\rm c}}{E}\right )^2}}\;\;.
\label{eq:osclength}
\end{equation}
Current constraints on $g_{\gamma a}$ for low-mass ALPs are set by the CAST experiment~\cite{2007JCAP...04..010A} and are of the order of $10^{-10}\;\rm GeV^{-1}$. For such couplings, and magnetic fields of a few $\mu$G, typical of clusters of galaxies~\cite{2002ARA&A..40..319C}, the critical energy lies at the TeV level for ALPs of mass $m_{\rm a}\sim 10^{-8}\;\rm eV$. For energies above $E_{\rm c}$, the strong mixing regime is attained. In the following, only ALP masses such that $E_{\rm c}$ is lower than the observation energy bands are considered. This defines the upper bound of the reachable ALP mass range.

In real astrophysical systems, the magnetic field is not homogeneous as considered in the previous example, but rather turbulent. In that case the propagation medium can be described as a collection of patches in which the magnetic field is oriented differently and have different values. The orientations and values of the magnetic field in the patches are related through the turbulence power spectrum, that can be considered for example as a Kolmogorov spectrum. The description of the beam can be done by the means of the density operator~\cite{Mirizzi:2009aj,2010JCAP...05..010B}
\begin{equation}
\rho\;\;=\;\;\left ( \begin{array}{c} A_1 \\ A_2 \\ a \end{array} \right ) \otimes \left ( \begin{array}{ccc} A_1 & A_2 & a \end{array} \right )^\star\;\;.
\end{equation}
In one domain, the mixing matrix $\mathcal{M}_k$ includes random values for the parameters $\phi$ and $B_t$ of Eq.~\ref{eq:matrix}. After $k$ domains, the composition of the beam is evaluated recursively as
\begin{equation}
\rho_{k+1} = e^{-i\mathcal{M}_k \,s}\cdot \rho_{k} \cdot e^{i\mathcal{M}^\dag_k \,s}\;\;,
\end{equation}
where $s$ is the size of the last domain. In the following the initial state is an unpolarized photon beam, described by $\rho_0=\text{diag}(\nicefrac{1}{2},\nicefrac{1}{2},0)$. The exact configuration of the magnetic field on the line of sight is unknown, so only the statistical properties of the overall transfer function can be predicted. In the case where $\lambda(E) \gg s$ in each domain, and for a large number $N$ of same-size domains, it can be shown~\cite{2002PhLB..543...23G} that the average overall transfer function is
\begin{equation}
P_{\gamma\rightarrow a}\;\;=\;\;\frac{1}{3} \left ( 1-\exp\left (- 3 \,N \,P_0 \right ) \right )\;\;,
\label{eq:aveffect}
\end{equation}
where $P_0$ is the transition probability for an unpolarized photon beam in one domain. At the transition between weak and strong mixing regimes, irregularities are expected to appear in the energy spectra of the sources~\cite{Wouters:2012qd}. This effect has been used to set constraints on low-mass ALPs, using H.E.S.S. data~\cite{HESS_ALP}, and Chandra data~\cite{2013arXiv1304.0989W}.

At this point it appears clearly that ALPs modify the propagation of photons in astrophysical media. For ALPs to affect the transparency of the Universe, magnetic fields such that the strong mixing regime is attained must be present on the line of sight. Those include magnetic fields in the source itself, in relation to the acceleration process of the high-energy particles. In case the source lies in a galaxy cluster, the cluster magnetic field can play a role. On larger scales the intergalactic magnetic field (IGMF) can be considered, and at the end of the photon journey, the galactic magnetic field (GMF) is relevant. The case of conversions in the IGMF has been widely studied, {\it e.g.} in~\cite{Mirizzi:2009aj}, the authors compute explicitly the mean transfer function and its variance. They show in particular that in general the variance of the transparency effect is large enough to include the conventional non-ALP case. It means that it can happen that the Universe is {\it less} transparent with ALPs than without. This is the case for instance if photons are converted to ALPs and do not convert back to photons before detection. It results that the effect can hardly be used to get constraints when IGMF only is considered. In addition, for the transparency to be significantly affected, the value of the IGMF that has to be assumed is of the order of 1 nG. That value is close to the current constraint~\cite{1999ApJ...514L..79B,2013arXiv1303.7121D} and should be considered as very optimistic. A safer scenario do not involve IGMF but rather magnetic fields close to the source and the GMF. This was proposed in~\cite{2008PhRvD..77f3001S} and then studied in more details in~\cite{Horns:2012kw}. In that case, $\gamma$ rays originate from the source and are converted into ALPs either in the source itself or in the surrounding galaxy cluster. Then, the ALP part of the beam travels unchanged from the source to the Milky Way, whereas the photons get absorbed on the EBL. When entering the Milky Way, the ALPs convert back into photons in the GMF. This can happen at high energies, where most photons are absorbed on the way, producing a potentially significant boost of the flux at high energy. 

\begin{figure}[h]
\centering
\includegraphics[width=0.5\textwidth]{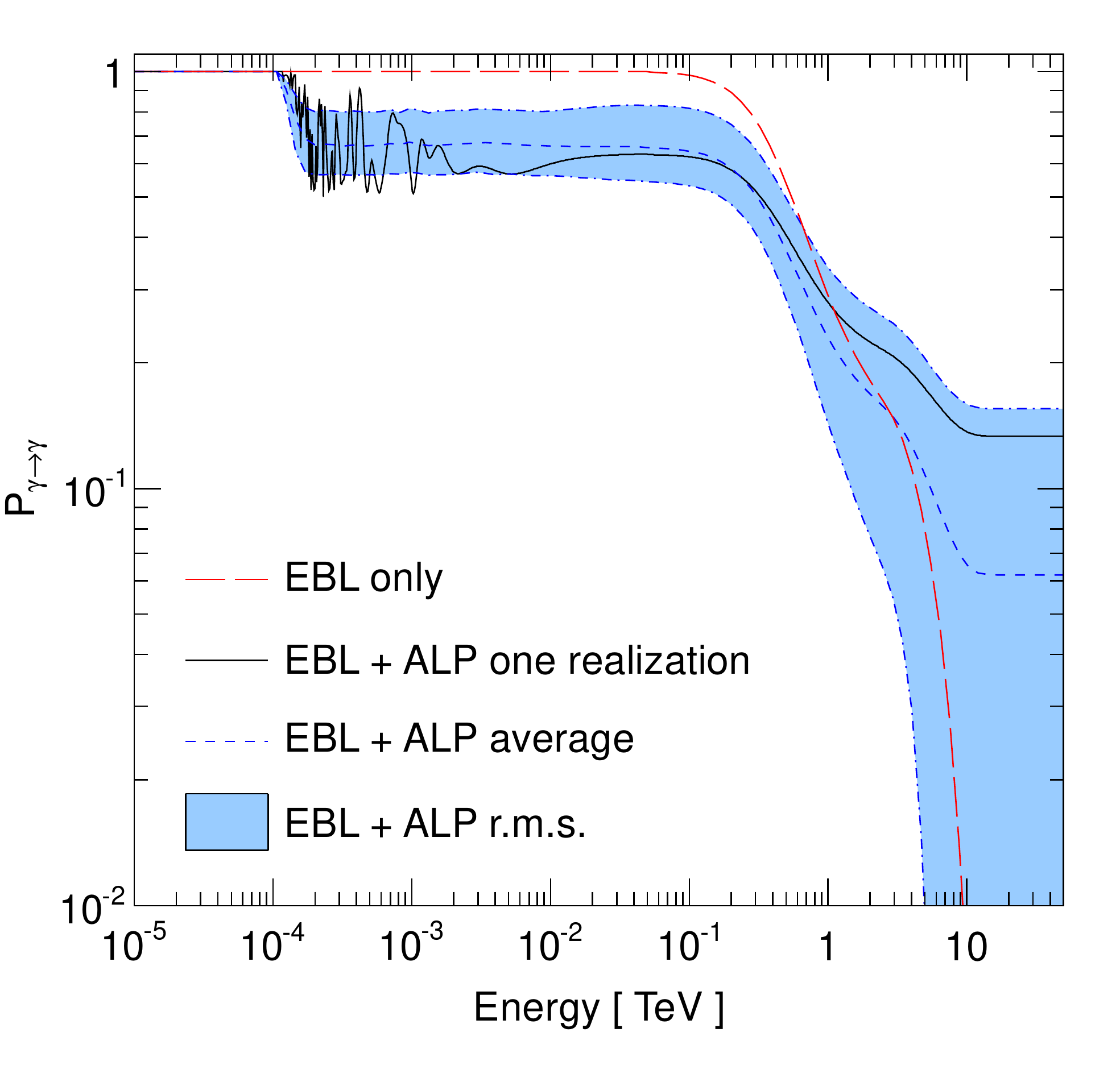}
\caption{Survival probability as a function of the energy. The dashed red line shows the conventional attenuation expected without ALPs. The dashed blue line shows the average over all realizations of the magnetic field while the blue band shows its variance. The solid black line is an example of one realization. An ALP with $g_{\gamma a}=5\times10^{-11}$ GeV$^{-1}$ and $m_{\rm a}= 1$ neV is considered. }
\label{fig:absorption}
\end{figure}

Figure~\ref{fig:absorption} gives an illustration of the overall transfer function for photons in such a scenario. For that figure, a source of VHE photons in a galaxy cluster located at $z=0.1$ is considered. The first conversions are assumed to occur in the galaxy cluster. In that generic case, a typical galaxy cluster magnetic field is considered, with a RMS value of $1\;\rm \mu G$ and a Kolmogorov power spectrum on scales from 10 kpc to 1 kpc. ALPs then convert back to photons in the GMF. On Fig.~\ref{fig:absorption}, the red-dashed line represents the transfer function in the case of EBL absorption only. Other curves are drawn for ALPs of mass $m_{\rm a}= 1$ neV and coupling $g_{\gamma a}=5\times10^{-11}$ GeV$^{-1}$. The blue-dashed curve is the mean value of the transfer function in the case of ALPs. One can see in particular that the opacity at high energy tends to be constant, in contrast with the conventional EBL case. The blue-shaded area is the 1-$\sigma$ envelope for this transfer function, corresponding to different possibilities of the magnetic field configuration on that specific line of sight. The solid black line is the transfer function in one randomly picked realization for the magnetic field. For that single observation, the irregularities appear around $E_{\rm c}$, and above 1 TeV the transparency becomes larger than in the conventional case. As shown in Fig.~\ref{fig:absorption}, the variance of the ALP scenario includes the conventional opacity prediction. It means that it is in principle impossible to derive constraints. Thus a statistical method based on the observation of a sample of sources is required. In the next section, the scenario is further motivated, and more details are given concerning the different conversion regions and the assumptions made in the subsequent analysis.

\section{The tested scenario}
\label{sec:2}

The bulk of extragalactic very high-energy $\gamma$-ray sources is composed of blazars.  These sources are also bright emitters of synchrotron radiation from radio to X rays, witnessing the presence of strong magnetic fields. They are therefore prime targets in the search for ALPs. In this study, $\gamma$/ALP oscillations in the source are assumed to be in the strong mixing regime. This assumption is verified if
\begin{equation}
g_{\gamma a}B L/2 \gtrsim 1\;\;,
\label{eq:strong_mixing}
\end{equation}
in natural units, or 
\begin{equation}
15\times\left (\frac{g_{\gamma a}}{{\rm 10^{-11} \;GeV^{-1}}}\right )\times\left ( \frac{B}{\rm G}\right )\times\left ( \frac{L}{\rm pc}\right ) \gtrsim 1\;\;,
\label{eq:strong_mixing}
\end{equation}
where $B$ is the magnetic field strength, and $L$ is the size of the conversion region~\cite{2007PhRvL..99w1102H}. For a value of the coupling strength allowed by current constraints, $g_{\gamma a} = 5\times10^{-11} \rm \;GeV^{-1}$, this condition translates to $B\times L \gtrsim 60 \, \rm \mu G\times kpc$. As shown in~\cite{2007PhRvL..99w1102H} this condition is similar to the Hillas criterion for the acceleration of ultra high energy cosmic rays, implying that the best candidates for the acceleration of UHECRs are also relevant targets to search for ALPs. These sources are typically active galaxy nuclei (AGN). Three different regions containing magnetic fields can be identified in the source. First, the magnetic field in the jets of the AGN, which are believed to host the processes responsible for the $\gamma$-ray emission, can be probed using multi-frequency radio observations~\cite{1998A&AS..132..261L,2005ApJ...619...73H}. From these observations, a magnetic field strength of a few mG on a scale of 10 pc is typically estimated in the jets~\cite{2008ASPC..386..451S,2009MNRAS.400...26O} of AGN. Second, the radio lobes at the extremities of the jets have magnetic fields of a few $\mu$G on scales of a few kpc~\cite{2005ApJ...626..733C}. Third, Fanaroff-Riley I radio galaxies, which are thought to be the parent population of BL-Lac objects, are frequently embedded in a galaxy cluster~\cite{1995ApJ...441..113S}. Typical magnetic fields in galaxy clusters are at the $\mu$G level on scales of a hundred of kpc~\cite{2002ARA&A..40..319C}. Blazars have their jets oriented close to the line of sight of the observer, so that the photons which are produced in these sources will cross all three domains. In each domain, the condition of strong mixing can be verified and globally, if one domain is skipped (the source may not be in a galaxy cluster for instance), the assumption of strong mixing in the source is still valid. In this study, the composition of the beam after propagation in the source is simulated following a random realization of a magnetic field of strength 1$\mu$G over a size of 500 kpc and turbulent on a scale of 10 kpc. As long as the strong mixing regime is attained, this simulation correctly describes the statistical distribution of the content of the beam, for any of the above mentioned conversion regions. Some statistical properties of this distribution, the mean and the variance, are formally derived in~\cite{Mirizzi:2009aj}. 

The beam of $\gamma$ rays and ALPs is propagated through the intergalactic medium. The EBL is modeled as in~\cite{2008A&A...487..837F}. $\gamma$ rays and ALPs may also mix toghether in the intergalactic magnetic field (IGMF). The nature and strength of the IGMF still remains unknown so that its modeling is hazardous. Current upper limits on the strength of the IGMF are at the level of 1 nG~\cite{1999ApJ...514L..79B,2013arXiv1303.7121D}, but for a significant mixing to occur over cosmological distances of Gpc scales, a strength of at least 0.1 nG is required following Eq. \ref{eq:strong_mixing} and still assuming $g_{\gamma a} = 5\times 10^{-11} \,\rm GeV^{-1}$. This means that a high value of the IGMF strength, close to the current upper limits, is needed to significantly mix photons and axions. Some studies even claim that the IGMF should be lower than $10^{-14}$ G to be in agreement with the high energy spectra of blazars~\cite{2013arXiv1303.5093F}. In the following, a zero IGMF is assumed, and the influence of a strong IGMF on the results is be discussed in Sec.~\ref{sec:7}.

\begin{figure}[h]
\centering
\includegraphics[width=0.8\textwidth]{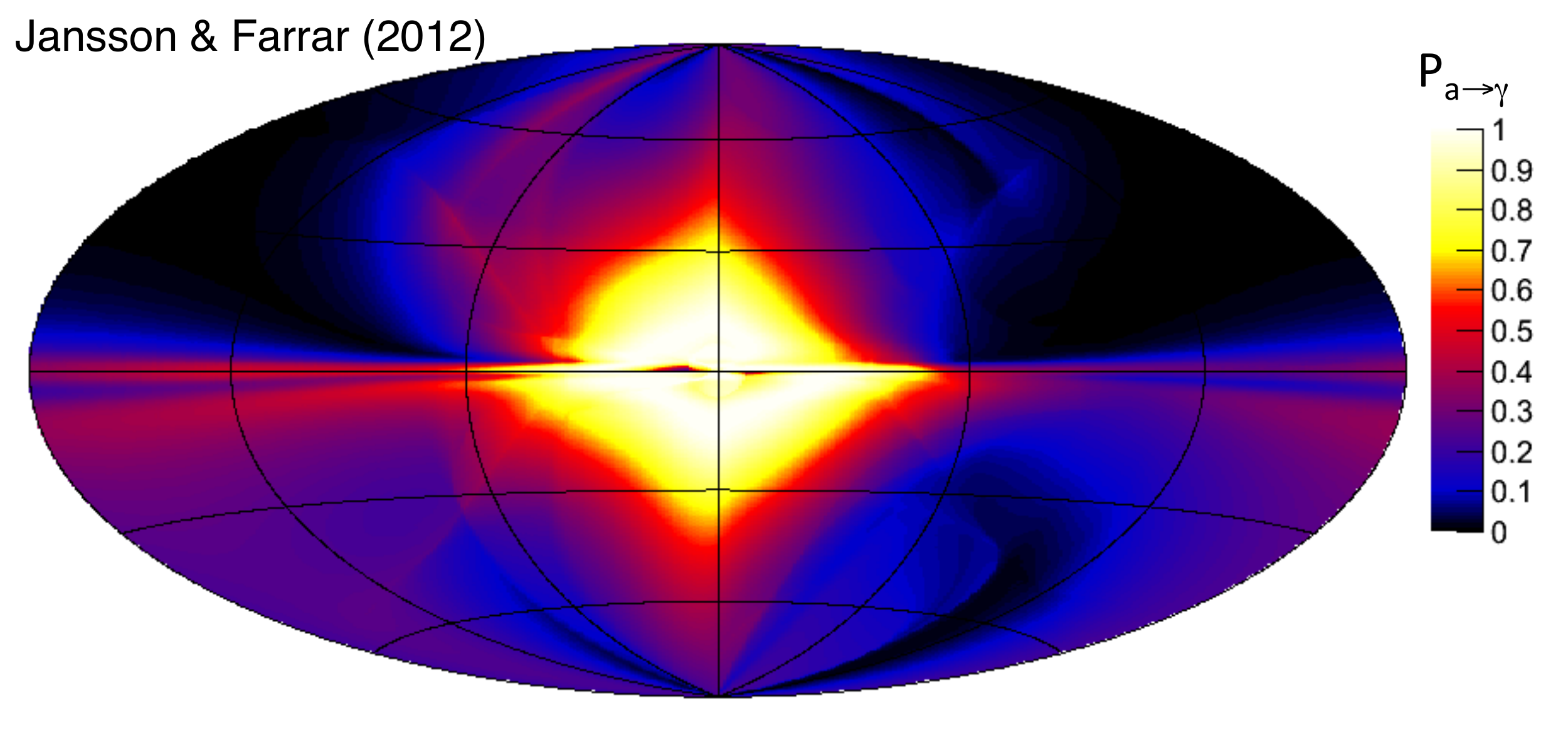}
\includegraphics[width=0.8\textwidth]{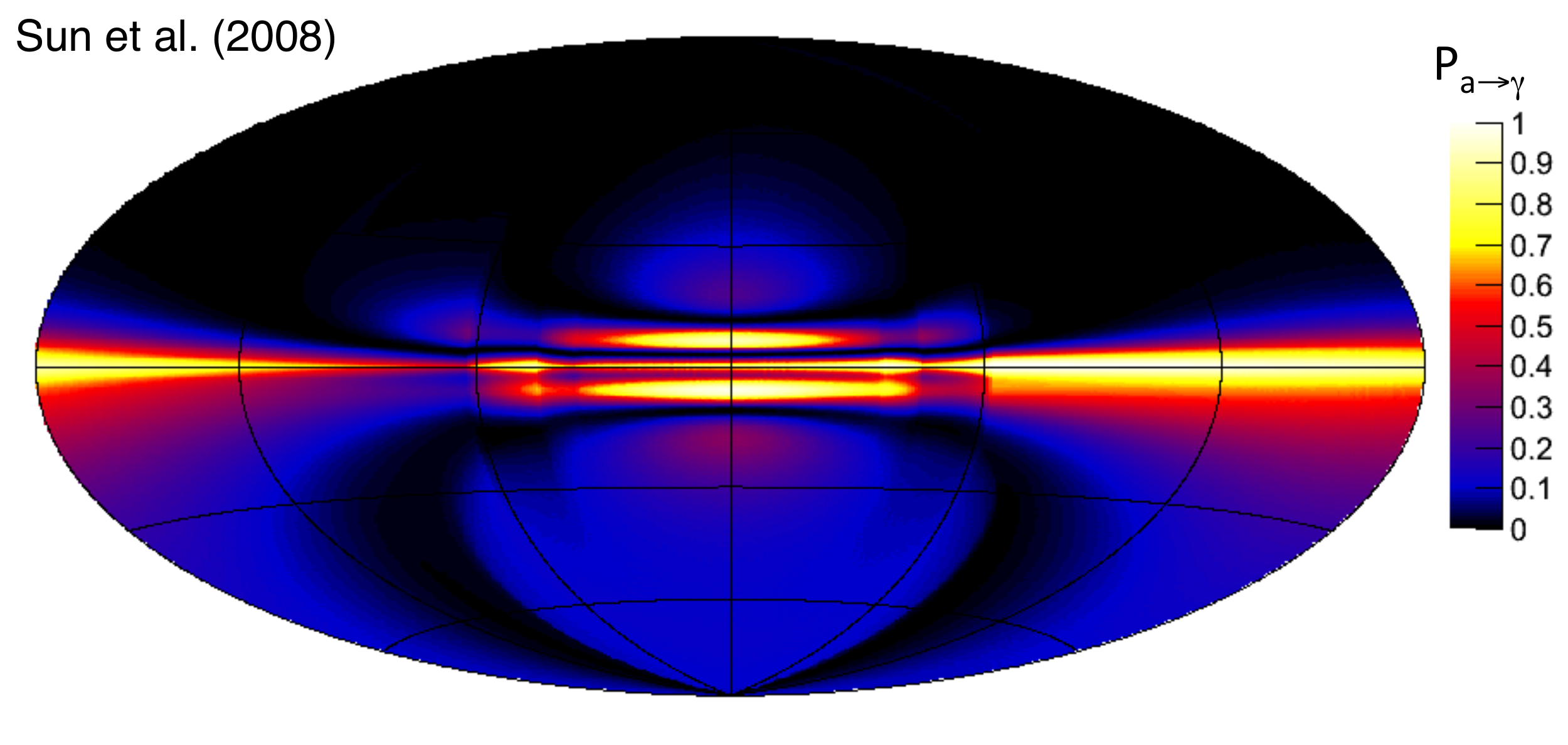}
\includegraphics[width=0.8\textwidth]{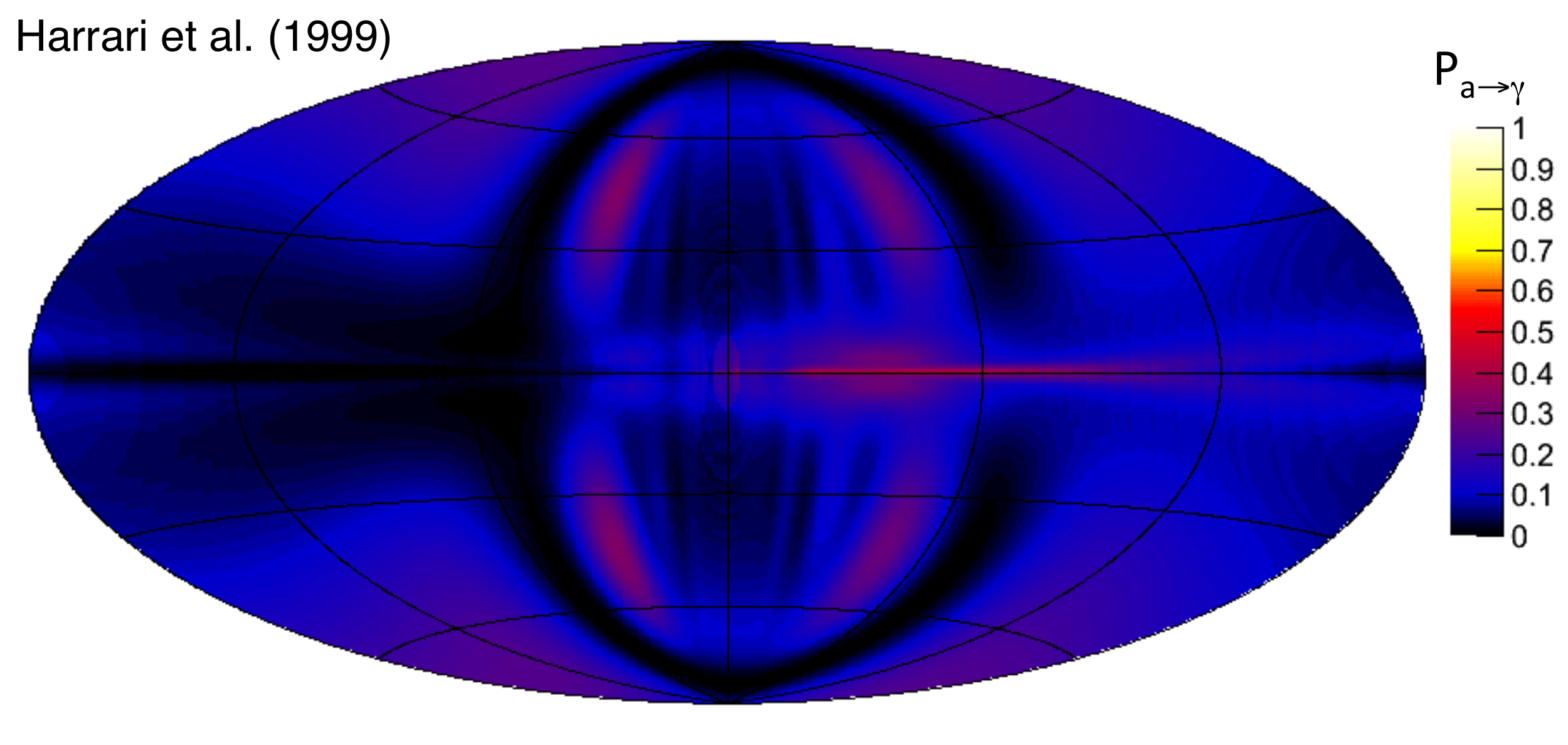}
\caption{Maps of probability of conversion from ALPs to photons in the galactic magnetic fields for three different models, \cite{2012ApJ...757...14J,2008A&A...477..573S,1999JHEP...08..022H} from top to bottom, assuming $g_{\gamma a} = 5\times10^{-11}$ GeV$^{-1}$. \label{fig:maps}}
\end{figure}

When entering the Milky Way, at energies corresponding to large $\tau$, the beam is mainly composed of ALPs since the photons have been absorbed on the EBL. During the propagation in the GMF, the ALPs can be back-converted into photons before reaching the Earth. The typical strength of the GMF is a few $\mu$G over more than 10 kpc, so that significant $\gamma$-ALP oscillations can take place. Here only the regular component of the GMF is of interest. The turbulent component is on average of the same order of magnitude than the regular one but is turbulent on small scales~\cite{1996ApJ...458..194M,2012ApJ...757...14J}. As shown in~\cite{2007PhRvD..76b3001M}, because of the large number of domains, the conversion does not happen efficiently for small turbulence scales, so that in the case of the GMF, the component turbulent on scales lower than 100 pc can be ignored (see also~\cite{Horns:2012kw}). In this study, three models for the regular component of the GMF are considered. The model from~\cite{1999JHEP...08..022H} is based on Faraday rotation measurements from pulsars and extragalactic sources, while the other two models~\cite{2008A&A...477..573S,2012ApJ...757...14J} also take into account the diffuse polarized synchrotron emission from plasma of relativistic electrons. For each direction on the sky, the conversion probability of ALPs to photons over the line of sight is computed using the full mixing matrix. A sky map of this probability is shown on Fig.~\ref{fig:maps} for the three models of the GMF. In the regions where this probability is the highest, the process is predicted to be more efficient and in these regions the Universe should appear to be the most transparent, {\it i.e.} the anomaly is expected to be there maximal.

In this scenario, the structures of the GMF are correlated with the measured hardening of the TeV spectra. If these structures were known to a high accuracy, it would be possible to use the cross-correlation between the measured spectral hardening  and the transverse magnetic field integrated on the same line of sight as a signature of this anomalous transparency scenario. However, as shown on Fig.~\ref{fig:maps}, the structures of the GMF for the three models considered here are not similar. More generally, no consensus is reached on the nature of the different components of the GMF and the resulting back-conversion map is model dependent. One common feature of these structures is that they show patterns that appear on similarly large angular scales. As a result, the hardening measured from two neighbor directions would appear correlated, whatever the GMF model. Conversely, two sources with large relative angular distance would appear anti-correlated. A model-independent and more robust signature of the scenario is thus provided by the angular autocorrelation of the  measured spectral hardening from various sources.

\section{The anisotropy test}
\label{sec:3}

\begin{figure}[h]
\centering
\includegraphics[width=0.5\textwidth]{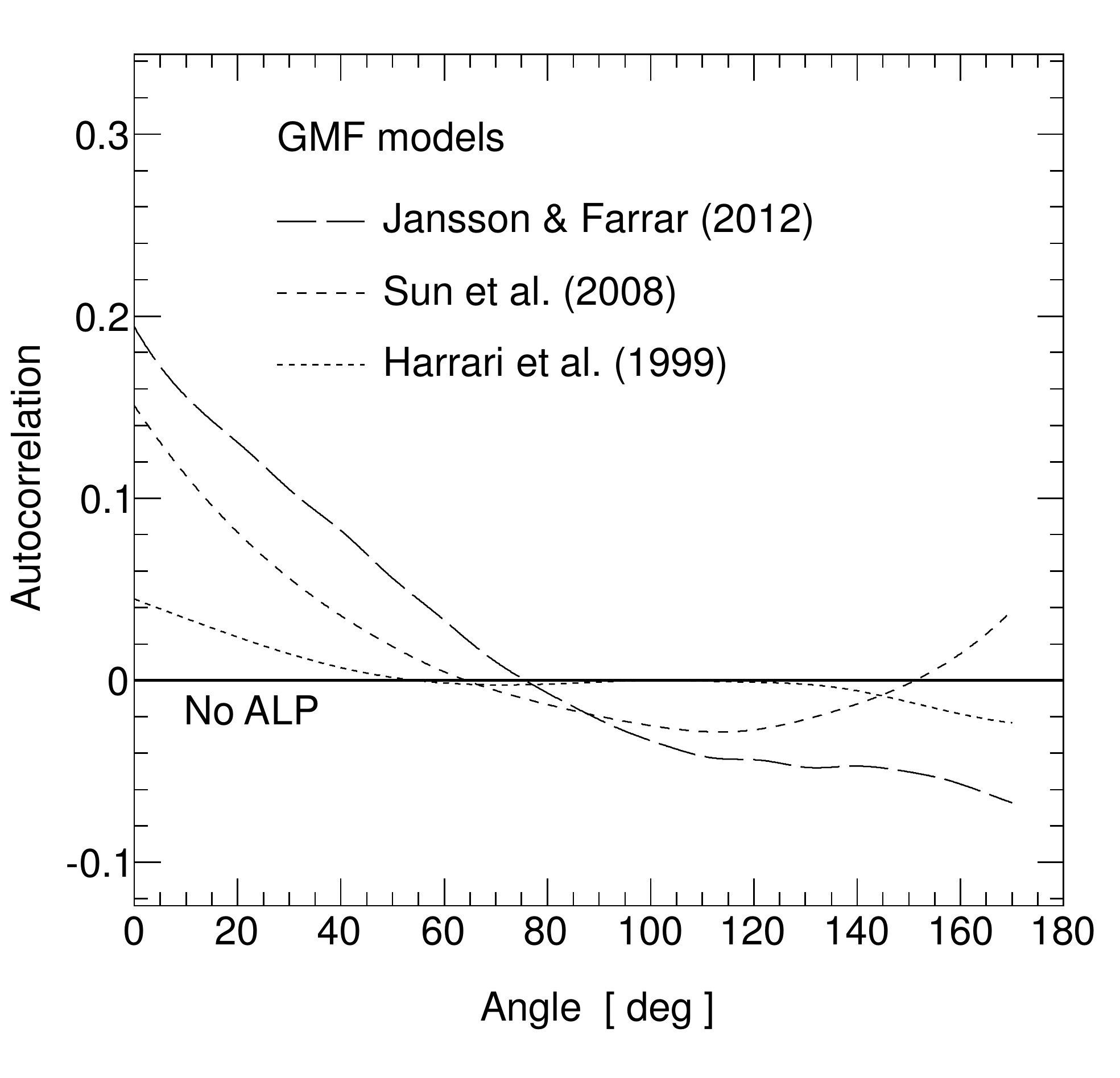}
\caption{Autocorrelation predictions with ALPs for the three considered GMF models.\label{fig:corr_GMF}}
\end{figure}

The spectral hardening is defined as $\Delta\Gamma = \Gamma_{\rm VHE} - \Gamma_{\rm HE}$ where $\Gamma_{\rm HE}$ and $\Gamma_{\rm VHE}$ are respectively the spectral indices measured in the high energy  (HE, 1 GeV < $E$ < 100 GeV) band and the very-high energy (VHE, E > 100 GeV) band. This observable has been introduced in~\cite{2013A&A...554A..75S} to study the effect of the EBL absorption in the conventional no-ALP case. The binned spatial autocorrelation of a sample of $N$ sources is computed in the following manner. In each bin of angular distance $\theta_i$ the autocorrelation $C_i$ is computed from the set of the $N_i$ couples of sources that are separated by an angular distance falling in the considered bin:
\begin{equation}
C_i = \frac{1}{N_i \, \sigma_{\Delta\Gamma_1}\,\sigma_{\Delta\Gamma_2}}\sum_{j=1}^{N_i}(\Delta\Gamma_1^j-\overline{\Delta\Gamma})(\Delta\Gamma_2^j-\overline{\Delta\Gamma}) \;\;,
\label{eq:corr}
\end{equation}
where $(\Delta\Gamma_1^j$ and $ \Delta\Gamma_2^j)$ are the spectral hardening values for the two sources of the $j$-th couple, and  $ \sigma_{\Delta\Gamma_1}\,\sigma_{\Delta\Gamma_2}$ are the rms of the $\Delta\Gamma$ distributions. The average spectral hardening over the whole sample is:
\begin{equation}
\overline{\Delta\Gamma} = \frac{1}{N}\sum_{i=1}^{N}\Delta\Gamma_i \;\;.
\label{eq:average}
\end{equation}
The theoretical autocorrelation expected from the scenario described in Sec.~\ref{sec:2} is shown on Fig.~\ref{fig:corr_GMF} for the three considered GMF models and $g_{\gamma a} = 5\times10^{-11} \rm \;GeV^{-1}$. It is computed by simulating a large number (5000) of sources randomly positioned on the sky and splining the autocorrelation calculated using Eq.~\ref{eq:corr} and \ref{eq:average}. The level of autocorrelation depends on the GMF model and on $g_{\gamma a}$ but the trends are similar for all models. At small angular distances, the spectral hardening values are correlated because the GMF structures are ordered on small scales. One consequence is that the spectral hardening appears to be slightly anti-correlated at the largest angular scales since the GMF structures will more likely be opposed when looking at scales much larger than their coherence scales. Note that the three models cross zero for quite similar angular distances, between $60^{\circ}$ and $80^{\circ}$. This is a common feature of the three considered GMF models.

The current sample of extragalactic VHE gamma-ray emitters, that will be described in the next section, is far from being in the limit of an infinite number of sources, so binned distributions are used. Moreover, the measurements of the spectral indices in the HE and VHE suffer from some uncertainties. These uncertainties result in a dispersion of the measured spectral hardening and in the end in a dispersion of the autocorrelation. The related uncertainty propagated on $C_i$ is then:
\begin{equation}
\sigma_{C_i}^2 = \frac{1}{N_i(N_i-1)}\sum_{j=1}^{N_i}\left(\frac{(\Delta\Gamma_1^j-\overline{\Delta\Gamma})(\Delta\Gamma_2^j-\overline{\Delta\Gamma})}{\sigma_{\Delta\Gamma_1}\,\sigma_{\Delta\Gamma_2}}-C_i\right)^2
\end{equation}
It has been checked that this procedure correctly implements the errors. In particular, fits of a constant performed over many realizations without ALP signal show that the statistics follow a correct $\chi^2$ distribution.

\section{Results with current data}
\label{sec:4}
\begin{figure}[h]
\centering
\includegraphics[width=0.8\textwidth]{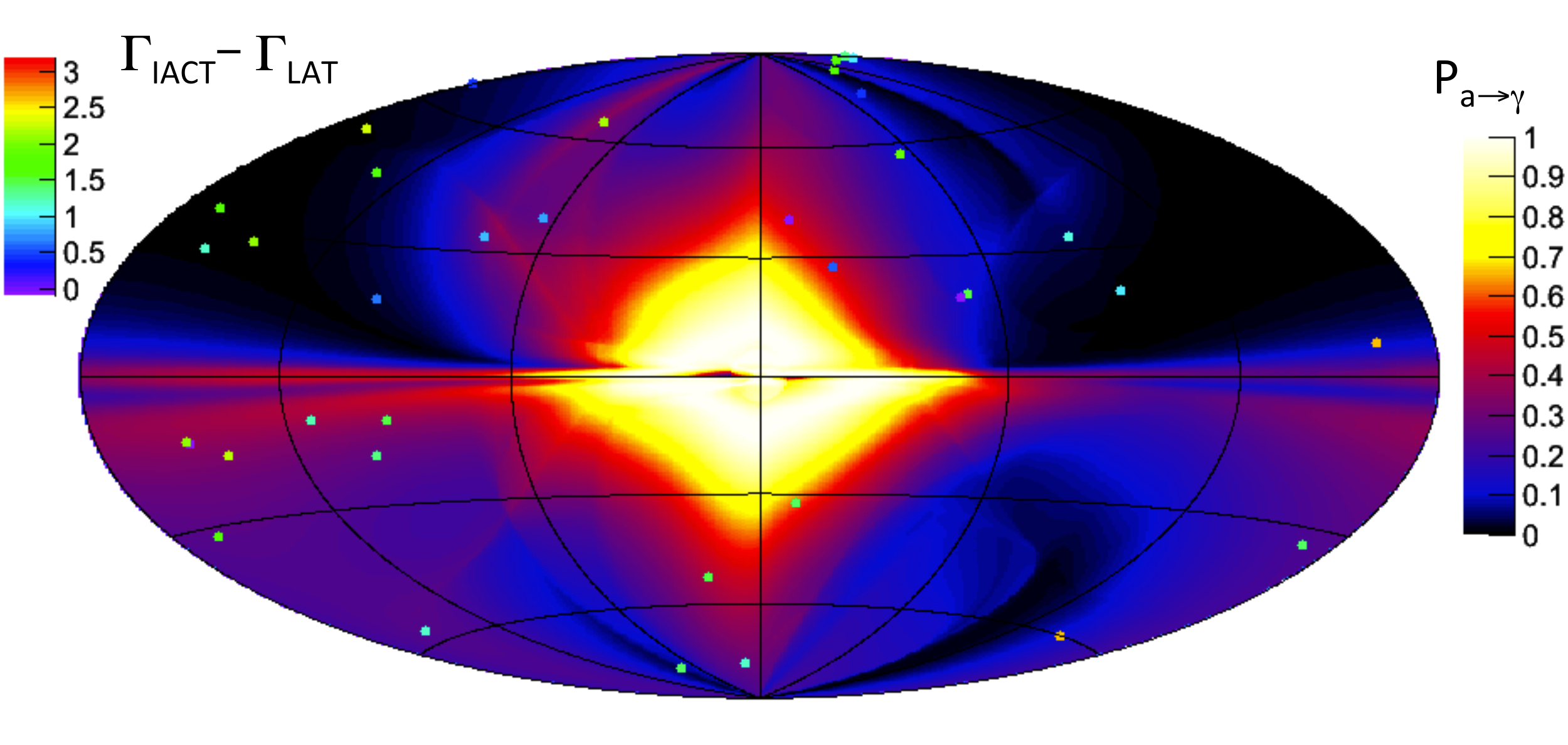}
\caption{Map of probability of conversion from ALPs to photons in the galactic magnetic field model of~\cite{2012ApJ...757...14J}. Overlaid is the spatial distribution of sources detected at VHE. The point colors indicate the break between spectral indices at VHE and HE.\label{fig:map_observed}}
\end{figure}

To date, 37 extragalactic sources are detected by \textit{Fermi}-LAT and IACTs with measured spectral indices in both bands and with known redshift. Most of those are BL-Lac objects with the exception of 4 radiogalaxies and 3 FSRQs. The list of the sources considered in this study is shown in Tab.~\ref{tab:sources}. The spatial distribution of these sources is shown on the sky map of Fig.~\ref{fig:map_observed} in galactic coordinates. The colors stand for the measured break between the spectral indices in the VHE and the HE band. The conversion probability of ALPs to photons in the GMF model of~\cite{2012ApJ...757...14J} is also shown.
\begin{table}
\begin{tabular}{cccccccc}
\hline\hline
Name 	&$l_{\rm J2000}$ (deg) & $b_{\rm J2000}$ (deg)  & redshift 	& Type 	& $\Gamma_{\rm LAT}$ & $\Gamma_{\rm IACT}$ & Ref.\\ \hline
SHBL 0013-185 	& 74.63	&	-78.09	&	0.095	& BL Lac	& 1.96 $\pm$ 0.20	&	3.4 $\pm$ 0.5 		& {\cite{2013A&A...554A..72H}}	\\
RGB J0152+017	& 152.37	&	-57.54	&	0.08		& BL Lac	& 1.79 $\pm$ 0.14	&	2.95 $\pm$ 0.36	& {\cite{2008A&A...481L.103A}}	\\
3C 66A			& 140.14	&	-16.77	&	0.44		& BL Lac	& 1.85 $\pm$ 0.02	&	3.64 $\pm$ 0.39	& {\cite{2011ApJ...726...58A}}		\\
IC 310			& 150.18	&	-13.73	&	0.0189	& RadG	& 2.10 $\pm$ 0.19	&	2.00 $\pm$ 0.14	& {\cite{2013arXiv1305.5147T}}	\\
PKS 0301-243		& 214.62	&	-60.18	&	0.266	& BL Lac	& 1.95 $\pm$ 0.05	&	4.6   $\pm$ 0.6		& {\cite{HESS_PKS0301}}\\
NGC 1275		& 150.58	&	-13.26	&	0.018	& RadG	& 2.00 $\pm$ 0.02	&	4.1	$\pm$ 0.7		& {\cite{2012A&A...539L...2A}}		\\	
RBS 0413			& 164.11	&	-31.70	&	0.19		& BL Lac	& 1.55 $\pm$ 0.11	&	3.18	$\pm$ 0.68	& {\cite{2012ApJ...750...94A}}		\\
1ES 0414+009		& 191.81	&	-33.16	&	0.287	& BL Lac	& 1.98 $\pm$ 0.16	&	3.45	$\pm$ 0.25	& {\cite{2012A&A...538A.103H}}	\\
RX J0648.7+1516	& 198.99	&	6.35		&	0.179	& BL Lac	& 1.74 $\pm$ 0.11	&	4.4	$\pm$ 0.8		& {\cite{2011ApJ...742..127A}}		\\
RGB J0710+591	& 157.41	&	25.43	&	0.125	& BL Lac	& 1.53 $\pm$ 0.12	&	2.69	$\pm$ 0.26	& {\cite{2010ApJ...715L..49A}}		\\
S5 0716+714		& 143.98	&	28.02	&	0.30		& BL Lac	& 2.00 $\pm$ 0.02	&	4.1	$\pm$ 0.7 	& {\cite{2009ApJ...704L.129A}}		\\
1ES 0806+524		& 166.26	&	32.91	&	0.138	& BL Lac	& 1.94 $\pm$ 0.06	&	3.6	$\pm$ 1.0		& {\cite{2009ApJ...690L.126A}}		\\
1RXS J1010-312	& 266.87	&	19.93	&	0.142	& BL Lac	& 2.09 $\pm$ 0.15	&	3.08	$\pm$ 0.42	& {\cite{2012A&A...542A..94H}}	\\
1ES 1011+496		& 165.53	&	52.73	&	0.212	& BL Lac	& 1.72 $\pm$ 0.04	&	4.0	$\pm$ 0.5		& {\cite{2007ApJ...667L..21A}}		\\
1ES 1101-232		& 273.17	&	33.06	&	0.186	& BL Lac	& 1.80 $\pm$ 0.21	&	2.94	$\pm$ 0.2~	& {\cite{2007A&A...470..475A}}		\\	
Mkn 421			& 179.82	&	65.04	&	0.031	& BL Lac	& 1.77 $\pm$ 0.01	&	2.20	$\pm$ 0.08	& {\cite{2007ApJ...663..125A}}		\\
Mkn 180			& 131.88	&	45.65	&	0.046	& BL Lac	& 1.74 $\pm$ 0.08	&	3.30	$\pm$ 0.7~	& {\cite{2006ApJ...648L.105A}}		\\
1ES 1215+303		& 188.93	&	82.06	&	0.13		& BL Lac	& 2.02 $\pm$ 0.04	&	2.96	$\pm$ 0.14	& {\cite{2012A&A...544A.142A}}	\\
1ES 1218+304		& 186.33	&	82.74	&	0.182	& BL Lac	& 1.71 $\pm$ 0.07	&	3.08	$\pm$ 0.34	& {\cite{2009ApJ...695.1370A}}		\\
W Comae			& 201.74	&	83.29	&	0.102	& BL Lac	& 2.02 $\pm$ 0.03	&	3.81	$\pm$ 0.35	& {\cite{2008ApJ...684L..73A}}		\\
4C +2135			& 255.07	&	81.66	&	0.432	& FSRQ	& 1.95 $\pm$ 0.21	&	3.75	$\pm$ 0.27	& {\cite{2011ApJ...730L...8A}}		\\
M 87				& 283.78	&	74.49	&	0.00436	& RadG	& 2.17 $\pm$ 0.07	&	2.60	$\pm$ 0.35	& {\cite{2006Sci...314.1424A}}		\\
3C 279			& 305.10	&	57.06	&	0.536	& FSRQ	& 2.22 $\pm$ 0.02	&	4.1	$\pm$ 0.7		& {\cite{2008Sci...320.1752M}}		\\
1ES 1312-423		& 307.55	&	20.05	&	0.105	& BL Lac	& 1.4   $\pm$ 0.4	&	2.9	$\pm$ 0.5		& {\cite{2013arXiv1306.3186H}}	\\
Centaurus A		& 309.52	&	19.42	&	0.00183	& RadG.	& 2.76 $\pm$ 0.05	&	2.7	$\pm$ 0.5		& {\cite{2009ApJ...695L..40A}}		\\
H 1426+428		& 77.472	&	64.89	&	0.129	& BL Lac	& 1.32 $\pm$ 0.12	&	~3.5	$\pm$ 0.35	& {\cite{2002ApJ...580..104P}}		\\
PKS 1510-089		& 351.29	&	40.14	&	0.361	& FSRQ	& 2.21 $\pm$ 0.03	&	5.4	$\pm$ 0.7		& {\cite{2013A&A...554A.107H}}	\\
AP Lib			& 340.68	&	27.58	&	0.049	& BL Lac	& 2.05 $\pm$ 0.04	&	2.5	$\pm$ 0.2		& {\cite{2011ICRC....8..107C}}		\\
Mkn 501			& 63.61	&	38.85	&	0.034	& BL Lac	& 1.74 $\pm$ 0.03	&	2.42	$\pm$ 0.2	~	& {\cite{2010A&A...524A..77A}}		\\
1ES 1727+502		& 77.12	&	33.55	&	0.055	& BL Lac	& 1.83 $\pm$ 0.13	&	2.7	$\pm$ 0.5		& {\cite{2013arXiv1302.6140M}}	\\
1ES 1959+650		& 98.02	&	17.67	&	0.048	& BL Lac	& 1.94 $\pm$ 0.03	&	2.58	$\pm$ 0.18	& {\cite{2008ApJ...679.1029T}}		\\
PKS 2005-489		& 350.37	&	-32.61	&	0.071	& BL Lac	& 1.78 $\pm$ 0.05	&	3.20	$\pm$ 0.16	& {\cite{2010A&A...511A..52H}}	\\
PKS 2155-304		& 17.74	&	-52.24	&	0.117	& BL Lac	& 1.84 $\pm$ 0.02	&	3.53	$\pm$ 0.06	& {\cite{2010A&A...520A..83H}}	\\
BL Lacertae		& 92.59	&	-10.44	&	0.069	& BL Lac	& 2.11 $\pm$ 0.04	&	3.6	$\pm$ 0.4		& {\cite{2013ApJ...762...92A}}		\\
B3 2247+381		& 98.25	&	-18.57	&	0.119	& BL Lac	& 1.83 $\pm$ 0.11	&	3.2	$\pm$ 0.5		& {\cite{2012A&A...539A.118A}}	\\
1ES 2344+514		& 112.89	&	-9.91	&	0.044	& BL Lac	& 1.73 $\pm$ 0.08	&	2.95	$\pm$ 0.12	& {\cite{2007ApJ...662..892A}}		\\
H 2356-309		& 12.84	&	-78.04	&	0.167	& BL Lac	& 1.89 $\pm$ 0.17	&	3.06	$\pm$ 0.24	& {\cite{2006A&A...455..461A}}		\\
\hline\hline
\end{tabular}
\caption{List of extragalactic sources detected by the \textit{Fermi}-LAT and IACTs with measured spectral indices and redshift.\label{tab:sources}}
\end{table}

\begin{figure*}[h]
\centering
\includegraphics[width=.49\textwidth]{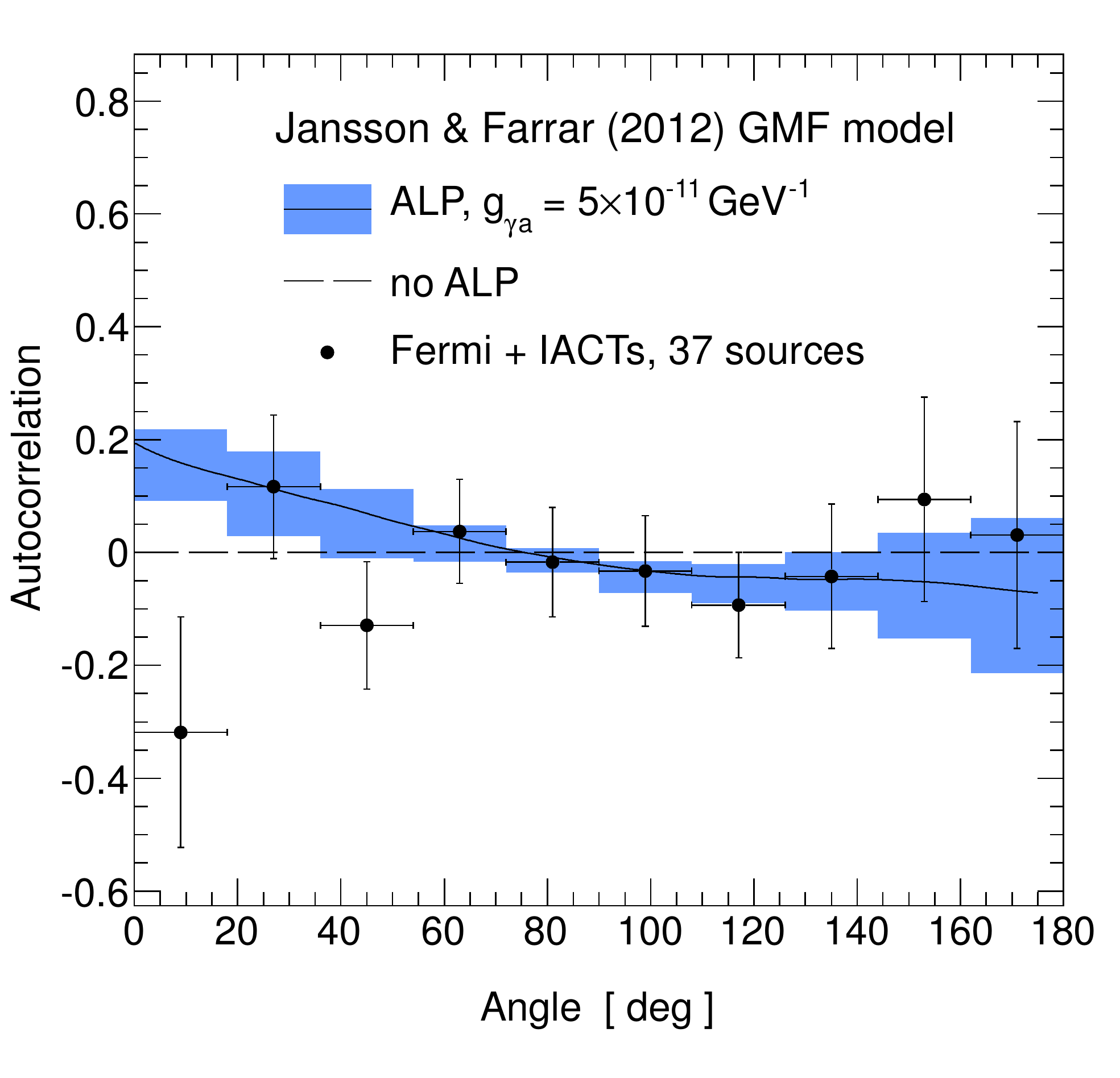}
\caption{Autocorrelation measured on the data , the plain black curve shows the model for $g_{\gamma a} = 5\times 10^{-11}$ GeV$^{-1}$ and the GMF of~\cite{2012ApJ...757...14J}. The blue boxes stand for the uncertainty on the model from the look-elsewhere effect (see text for details). \label{fig:observed}}
\end{figure*}

The spatial autocorrelation pattern for this set of sources is computed using \textit{Fermi}-LAT measurements of the spectral index from~\cite{2012ApJS..199...31N} in the HE band and measurements by IACTs in the VHE band. The measured autocorrelation is shown in Fig.~\ref{fig:observed}. The measurements are compatible with the no-autocorrelation hypothesis with a probability of 0.80. It has been checked that the results do not depend on the choice of the binning. The measurements seem to disfavor the autocorrelation model with an ALP coupling of $5\times10^{-11}$ GeV$^{-1}$ and the GMF model of~\cite{2012ApJ...757...14J} , also shown in Fig.~\ref{fig:observed}. However, the current list of detected sources is still rather limited. This implies a limited sampling of the sky, and an intrinsic randomness in the observation of the ALP effect. It  could be for instance that the sources are all located in areas of similar back-conversion probability, leading to an observed transparency effect apparently independent of the position of the sources. This would translate into an autocorrelation pattern compatible with zero. To account for the trial factors for the look-elsewhere effect, simulations are performed where the source positions are scrambled. This allows to estimate the uncertainty that arises from their configuration, for which only one realization is observed. The 1-$\sigma$ uncertainty on the model from the look-elsewhere effect is shown in Fig.~\ref{fig:observed} as blue boxes. When this effect is taken into account, the observed dataset does not allow to disentangle the ALP transparency model from the no-ALP case. It is found that the probability of the ALP model is 0.7 taking into account the look-elsewhere effect. Given the data points, the probability of the no-autocorrelation hypothesis is 0.8. This does not allow to confirm or reject the ALP hypothesis. Current data cannot exclude the ALP case because the number of detected sources is too small. In the future, the CTA project will enable the discovery of numerous extragalactic sources, thus increasing the sensitivity to the autocorrelation observable.

\section{Expected sample of extragalactic sources with CTA}
\label{sec:5}

CTA is the next-generation array of Cherenkov telescopes. It will consist of two large arrays of tens of telescopes of different sizes, the configuration of which are not yet fully defined~\cite{Consortium:2010bc}. Two CTA sites are foreseen to allow a survey of the whole sky, one in the southern hemisphere and one in the northern one. In the following we assume an even coverage of the whole sky, which can be achieved with a long enough operation of CTA in survey mode~\cite{Dubus:2012hm}. The detectability of AGN with CTA has been thoroughly discussed, for a recent study, see~\cite{2013APh....43..215S}. In~\cite{2013APh....43..215S}, it is estimated that at least 500 AGN will be detected by CTA above 30 GeV in ten years in the array B configuration (see Fig.~7 of~\cite{2013APh....43..215S}), which is optimized for low energies,~\cite{Consortium:2010bc}. In the following, it is assumed that 500 extragalactic sources with known redshifts are detected by CTA. In Sec.~\ref{sec:7}, the consideration of other array configurations is briefly discussed. Simulations of this sample of sources are performed to estimate the sensitivity to the angular autocorrelation observable that could be achieved with CTA. The cumulative distribution function (CDF) for redshifts of the blazars (BL-Lac objects and FSRQs) that should be detected by CTA in the array B configuration is shown in Fig.~7 of~\cite{2013APh....43..215S}. This distribution is obtained by simulating spectral energy distributions of the sources based on a blazar $\gamma$-ray luminosity function in agreement with \textit{Fermi}-LAT results and with conservative assumptions regarding the standard blazar sequence. This prior CDF of the redshift is used to randomly generate the redshifts of the simulated source sample. 

The extragalactic sky will mostly be observed by CTA in survey mode. The distribution of the detection significance of the sources in the sample should then closely match the distribution found in~\cite{2012ApJS..199...31N} for the sample of sources detected by \textit{Fermi}-LAT which is also operated in survey mode. In this distribution, shown in Fig.~23 of~\cite{2012ApJS..199...31N}, the bulk of detected sources has a significance close to $5~\sigma$, the threshold for a detection. The distribution then rapidly decreases as brighter sources are scarcer. To each source in the simulated CTA sample, a significance is randomly associated with a prior following the \textit{Fermi}-LAT in the 2 year catalog (2FGL) distribution of source significance. This detection significance is then converted into a number of events in the source region (ON region) using the Eq. 17 of~\cite{1983ApJ...272..317L} and an estimation of the expected background that would be measured in sample regions around the ON region (OFF regions). The estimation of the background is given in the Monte-Carlo study for CTA of~\cite{2013APh....43..171B}. The number of excess events associated to the source, that are used for the spectrum reconstruction, is eventually obtained by subtracting the number of events in the ON region by the number of events in the OFF regions normalized by the respective region areas.

The spectrum of the excess events is built following a power law with absorption on the EBL and transparency effect due to ALPs. The shape is the following:
\begin{equation}
\phi \propto \left (\frac{E}{E_0}\right )^{-\Gamma}\left[P_{\gamma\rightarrow a}^{~\rm src}P_{a\rightarrow\gamma}^{~\rm MW} + (1-P_{\gamma\rightarrow a}^{~\rm src})e^{-\tau_{\gamma\gamma}(E)}P_{\gamma\rightarrow\gamma}^{~\rm MW}\right] \;\;.
\end{equation}
The first term inside the brackets corresponds to the flux of photons converted into ALPs in the source that are back-converted to photons in the Milky Way before detection. The second term is the flux of photons that are not converted in the source and get absorbed in the intergalactic medium. The slope of the power law $\Gamma$ is randomly generated following the distributions of spectral indices measured by \textit{Fermi}-LAT in the 2FGL and shown in Fig.~21 of~\cite{2012ApJS..199...31N}. The relevance of spectral breaks intrinsic to the source is discussed in Sec.~\ref{sec:7}. The excess events are distributed following this spectral shape and convolved with the effective area expected for CTA given in~\cite{2013APh....43..171B}. The energy distribution of the excess events is then binned in energy bins of relative size of 10\% of the energy. This width is typical of the energy resolution that can be achieved with CTA~\cite{2013APh....43..171B}. The simulated spectrum is eventually obtained by dividing the histogram by the effective area. A re-binning procedure ensures a significance of at least 2 $\sigma$ per point. A mock spectrum, as it would be observed by CTA, is built through this procedure for each source. Figure~\ref{fig:cta_spec} shows two examples of would-be measured spectra for sources detected a the 50-$\sigma$ and 5-$\sigma$ level, and redshifts of $z=0.1$ and $z=0.7$ respectively. The simulated sample of sources is then used to estimate the sensitivity of CTA to the spatial autocorrelation of spectral hardening values.

\begin{figure}[h]
\centering
\includegraphics[width=0.8\textwidth]{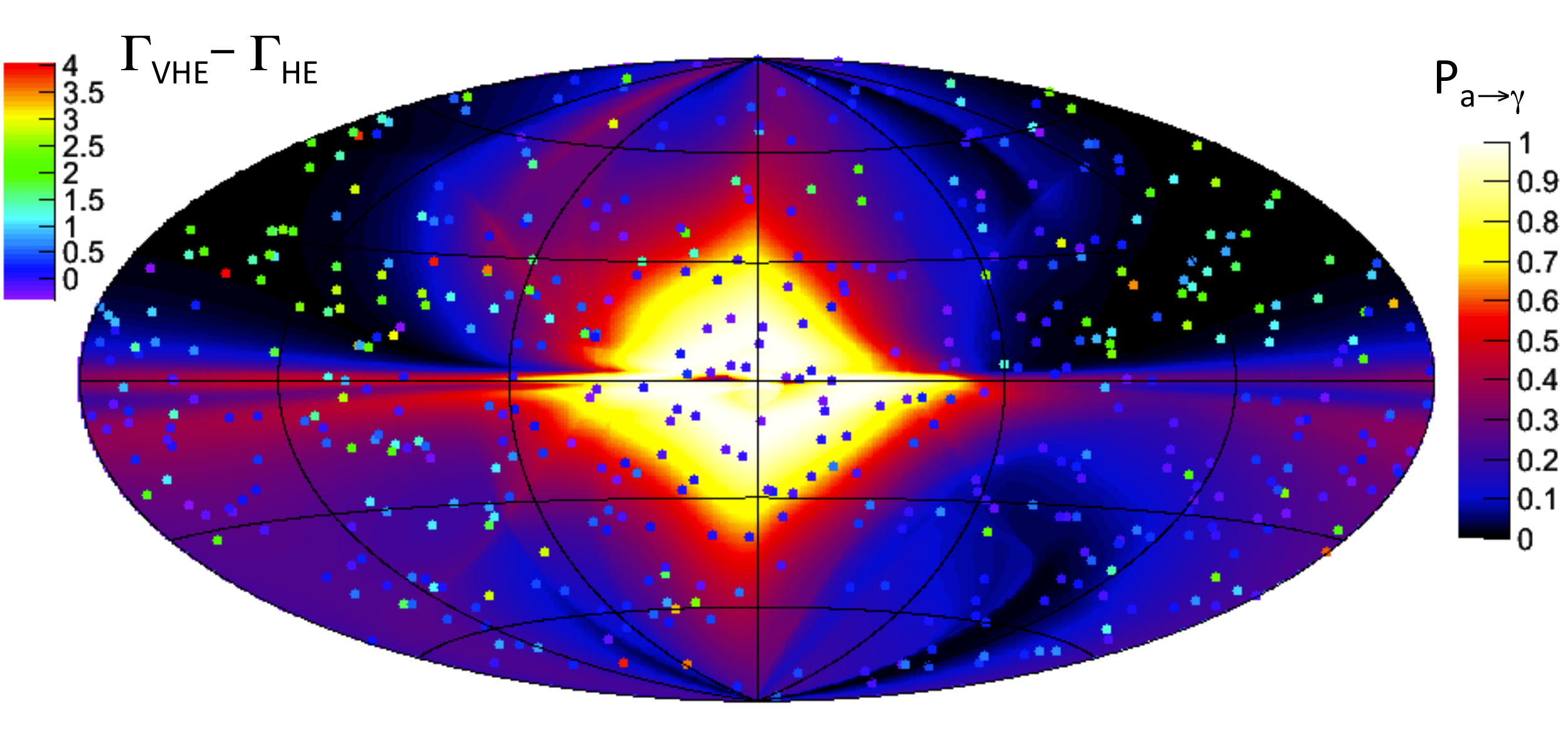}
\caption{Map of probability of conversion from ALPs to $\gamma$ rays in the galactic magnetic field model of~\cite{2012ApJ...757...14J}. Overlaid is the spatial distribution of a simulated sample of 500 sources that could be detected by CTA, assuming $g_{\gamma a} = 5\times 10^{-11}$ GeV$^{-1}$. The point colors indicate the break between spectral indices at VHE and HE. \label{fig:simu}}
\end{figure}

\begin{figure}[h]
\centering
\includegraphics[width=0.49\textwidth]{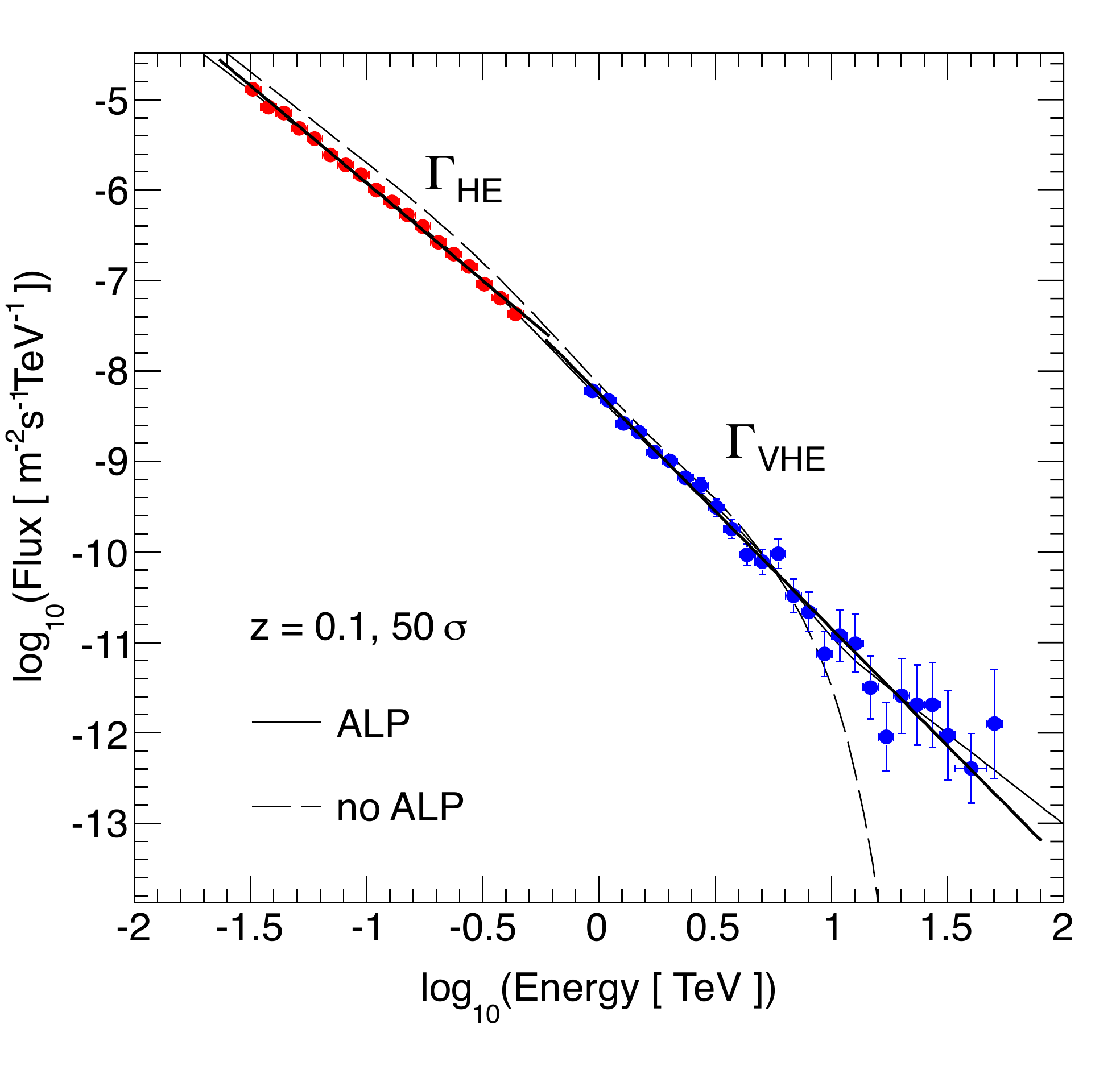}
\includegraphics[width=0.49\textwidth]{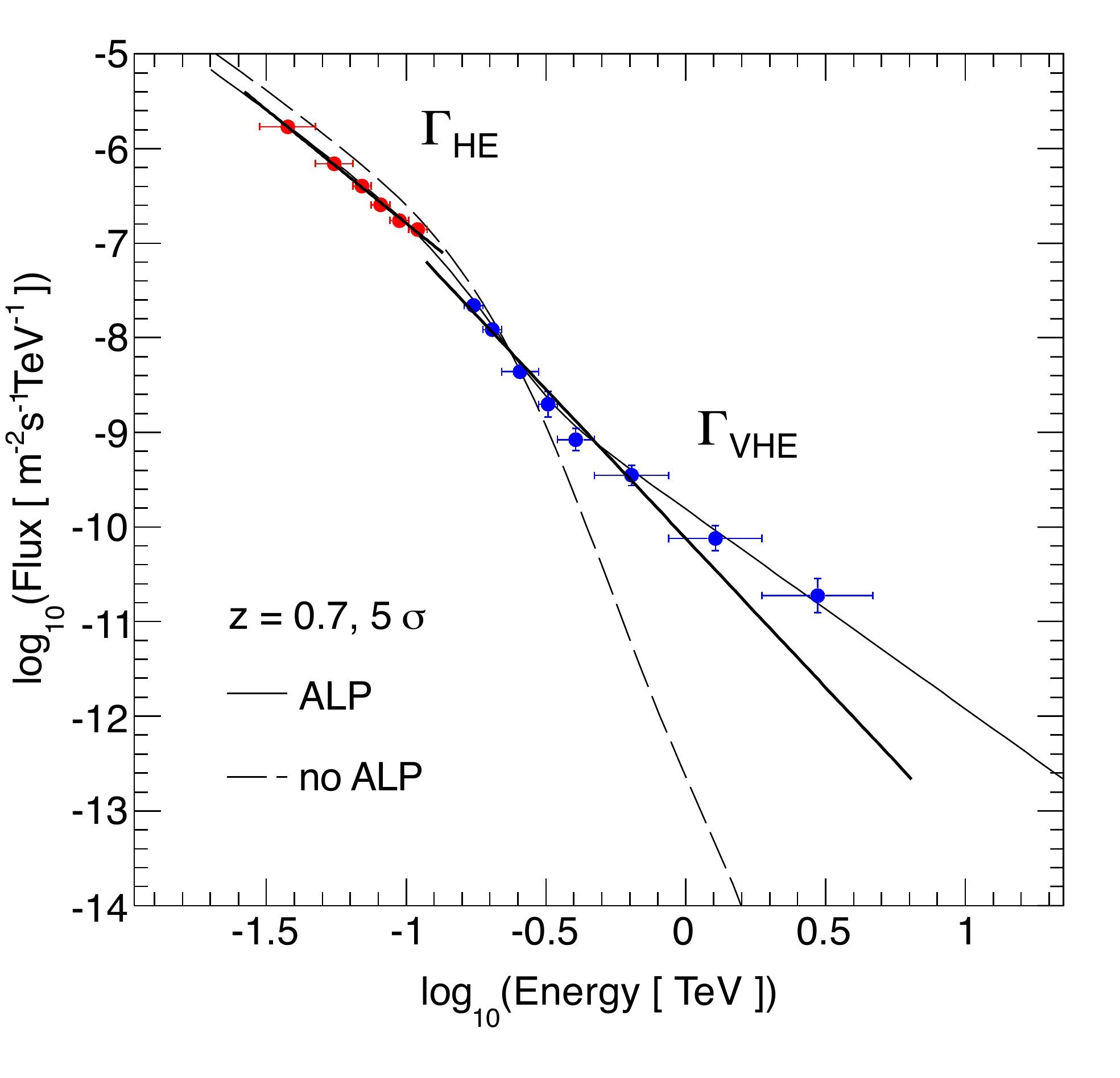}
\caption{Examples of two sources spectra as they would be observed by CTA, for sources located at $z=0.1$ and $z=0.7$, with significance of detections of 50 $\sigma$ and 5 $\sigma$ for the left and right panels respectively. Simulated observations correspond to the ALP scenario with $g_{\gamma a} = 5\times10^{-11}$ GeV$^{-1}$.\label{fig:cta_spec}}
\end{figure}

\section{CTA sensitivity to ALPs using angular correlations}
\label{sec:6}

\begin{figure}[h]
\centering
\includegraphics[width=.49\textwidth]{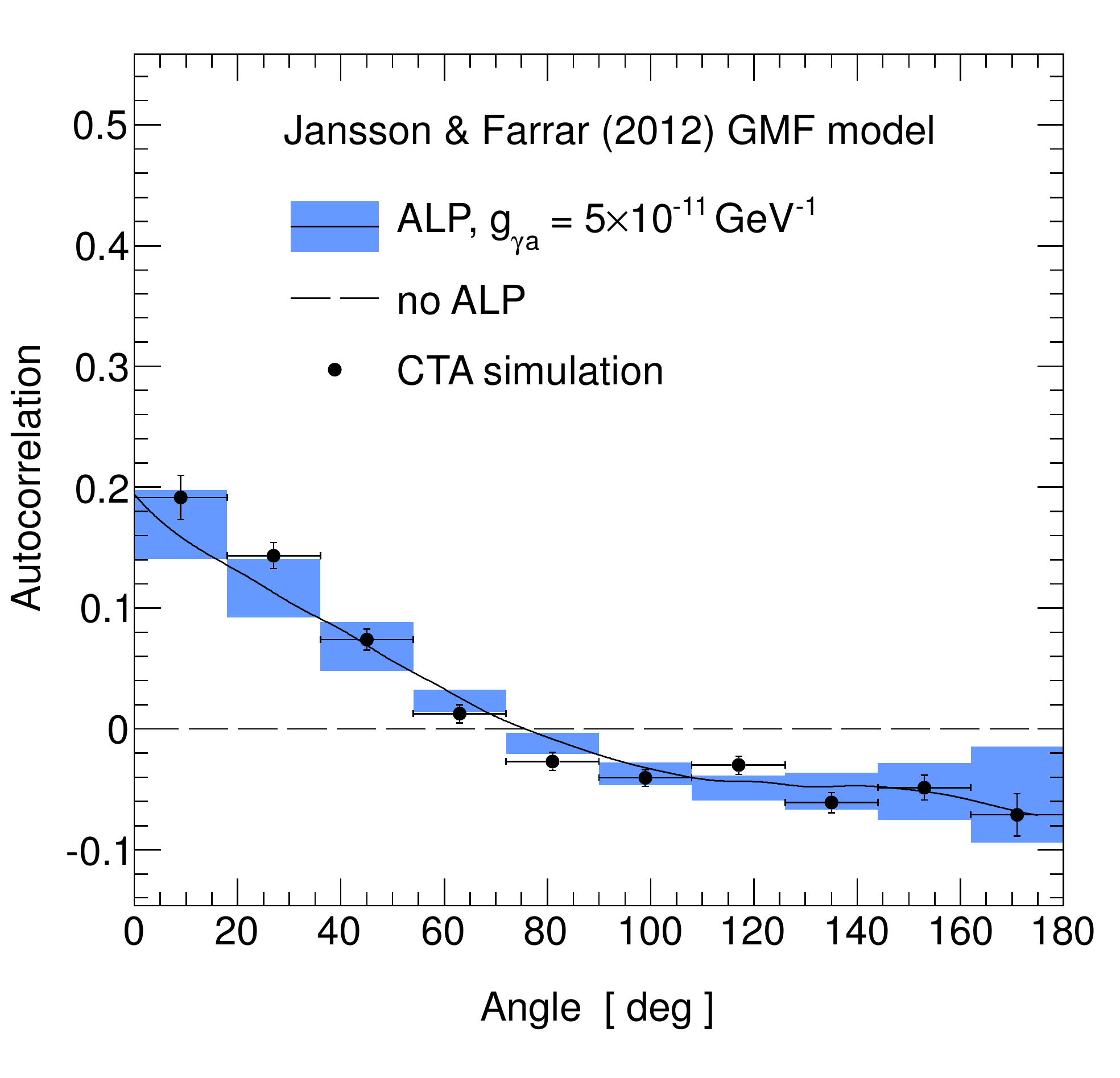}
\caption{Autocorrelation measured on a simulated sample of 500 sources that could be detected by CTA with $g_{\gamma a} = 5\times 10^{-11}$ GeV$^{-1}$ and the GMF of~\cite{2012ApJ...757...14J}. The plain black curve shows the ALP prediction for the same coupling strength and GMF model. The blue boxes stand for the uncertainty on the model from the look-elsewhere effect (see text for details). \label{fig:corr_simu}}
\end{figure}

By covering a large energy range, CTA will be able to measure simultaneously the spectral index in the HE band, not affected by EBL absorption, and the spectral index in the VHE band. Here it is assumed that the redshift is known for all considered sources. For each source, the energy separating the two domains is given by the optical depth of absorption on the EBL without ALP-induced transparency effects. The spectral index in the HE band is measured on energy bins for which the optical depth is lower than 0.5.  This low value ensures that the spectral index measured in this energy band is not affected by EBL absorption. Conversely, for the VHE band, energy bins with optical depth greater than 1 are used, so that the measured slope will be effectively affected by the absorption on the EBL, and possibly modified by ALP effects. The corresponding energy bands for the two examples of Fig.~\ref{fig:cta_spec} are shown in the corresponding plots. A sky map of the spectral hardening values measured in one set of 500 simulated CTA source observations is shown in Fig.~\ref{fig:simu} for the GMF model of~\cite{2012ApJ...757...14J}. In the example of Fig.~\ref{fig:simu}, ALP effects are considered and simulations are performed assuming $g_{\gamma a} = 5\times10^{-11}\rm~GeV^{-1}$. A correlation between the spectral hardening and the ALP back-conversion in the Milky Way probability is visible through the color codes. The autocorrelation measured for the simulation of this set of sources is shown in Fig.~\ref{fig:corr_simu}. As in Fig.~\ref{fig:observed}, the blue boxes stand for the theoretical uncertainty from the look-elsewhere effect. Due to the much larger number of sources as compared to the currently detected sample, the uncertainty on the points is reduced, as well as the uncertainty arising from the look-elsewhere effect. When combining these two uncertainties, the probability of the no-autocorrelation hypothesis for the single displayed realization is $8\times10^{-20}$, this corresponds to a 8.4-$\sigma$ deviation. This sample of simulated sources clearly exclude the no-autocorrelation hypothesis, showing that CTA will be sensitive to the autocorrelation observable.

Figures~\ref{fig:simu} and \ref{fig:corr_simu} are shown for a coupling constant of $g_{\gamma a} = 5\times10^{-11}\rm~GeV^{-1}$. When decreasing $g_{\gamma a}$, the measured autocorrelation is more compatible with the no-ALP hypothesis. The sensitivity of CTA is computed by estimating the lowest coupling strength that could be excluded assuming that no significant autocorrelation is observed. To estimate this sensitivity, a large number of sample of sources (1000) are simulated with $g_{\gamma a} = 0$, assuming that no signal is observed. From the corresponding autocorrelation distributions, predicted shapes for the autocorrelation are fitted with increasing $g_{\gamma a}$ until the coupling is excluded at the 95\% C.L. For each generated sample, the exclusion set on $g_{\gamma a}$ is different because of the different source positions in the realizations. The averaged exclusion over all the generated sample at the 95\% C.L. is $g_{\gamma a} < 2.92\times10^{-11}\rm~GeV^{-1}$. The variance of the level of exclusion over the whole set of realizations is $2.9\times10^{-12}\rm~GeV^{-1}$.

\begin{figure}[h]
\centering
\includegraphics[width=0.6\textwidth]{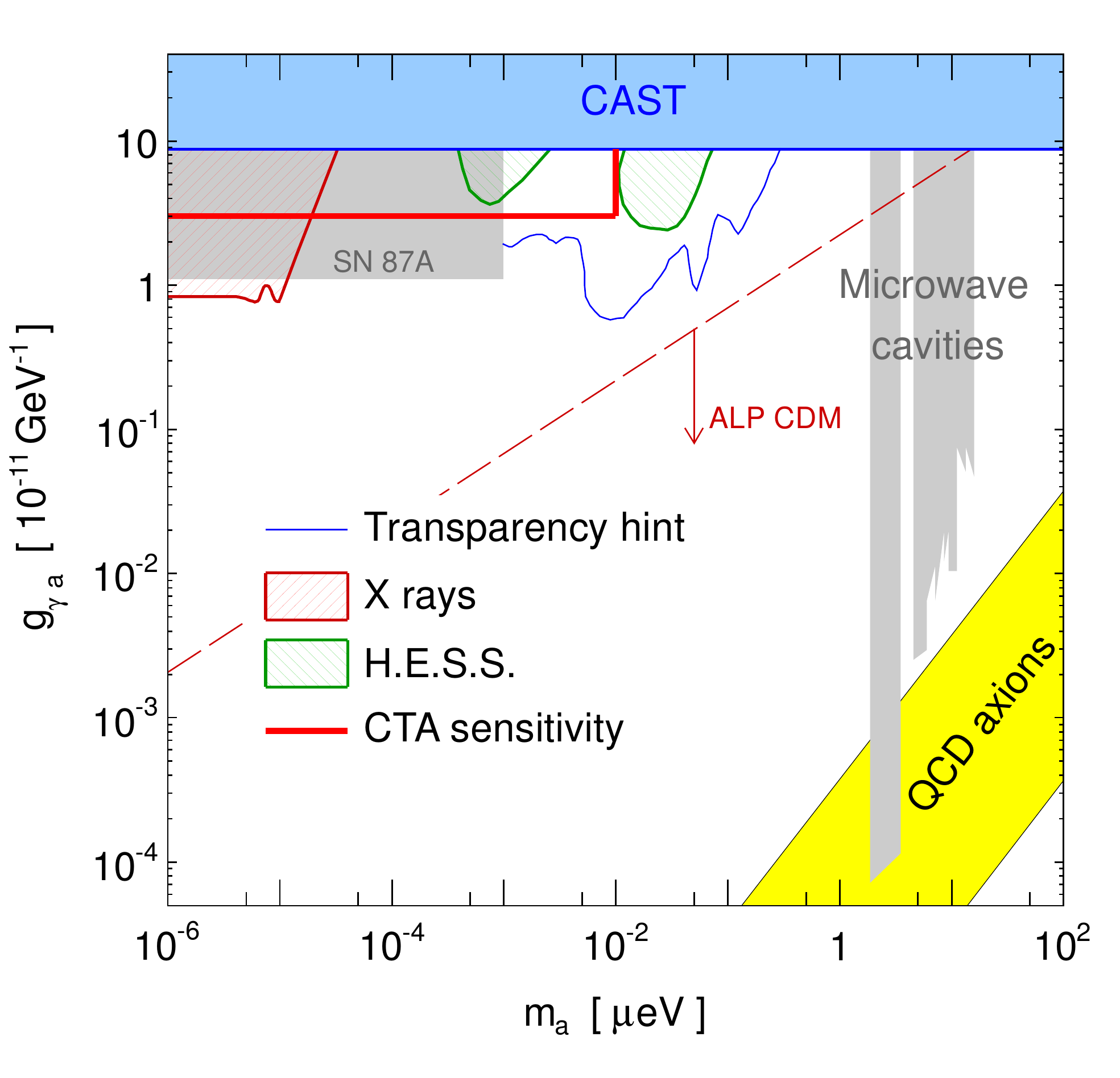}
\caption{ALP parameter space showing the sensitivity of CTA with the autocorrelation observable. Also shown are various constraints in the same mass range (see Sec.~\ref{sec:intro} for details). The range of ALP parameters that could explained the opacity anomaly is also shown as "Transparency hint"~\cite{2013PhRvD..87c5027M}. \label{fig:sensitivity}}
\end{figure}

As explained in Sec.~\ref{sec:2}, when considering $\mu$G level magnetic fields, like the magnetic fields in the source and in the Milky Way, the coupling between ALPs and photons is efficient above tens of GeV for ALP masses lower than 10 neV. The corresponding range of ALP parameters accessible with CTA is shown in Fig.~\ref{fig:sensitivity}. On this figure, other constraints from independent analyses and experiments are also shown. For higher masses, the $\gamma$-ALP conversion is not in the strong regime and spectral irregularities occur. Current exclusions might be extended by looking at the regularity of the spectrum of bright TeV sources with CTA, which is beyond the scope of the present study.

\section{Discussion}
\label{sec:7}

One possible limitation of the test is that the redshifts of all the sources may not be known. In the 2FGL, two third of the sources have known redshift, which means that one can expect one third of the prospected CTA sample to not have a measurement of the redshift. Knowledge of the redshift of the source is not crucial to the test and it can be worked around if no measurements are available. The redshift of the source is used in the analysis here to define the energy ranges of the HE and VHE domain. For example in Sec.~\ref{sec:4}, the energy intervals are defined from whether {\it Fermi}-LAT or an IACT are used to determine the spectral index, the redshift of the sources are not used at that point. In Sec.~\ref{sec:5}, the ranges are defined as a function of the optical depth, for which the redshift of the source is required. If the redshift of the source is not known, some other definitions for the HE and VHE ranges can then be used. With CTA, the analysis could be performed with fixed energy intervals. The consequence might be to broaden the $\Delta\Gamma$ distributions, but as long as no spatial correlation are introduced, the autocorrelation observable is still efficient. One can even think of other methods to define these intervals, for instance, the numerical derivative calculated on the spectral points could be used to determine when there is a significant curvature, that would mark the transition between both domains.

\begin{figure}[h]
\centering
\includegraphics[width=0.5\textwidth]{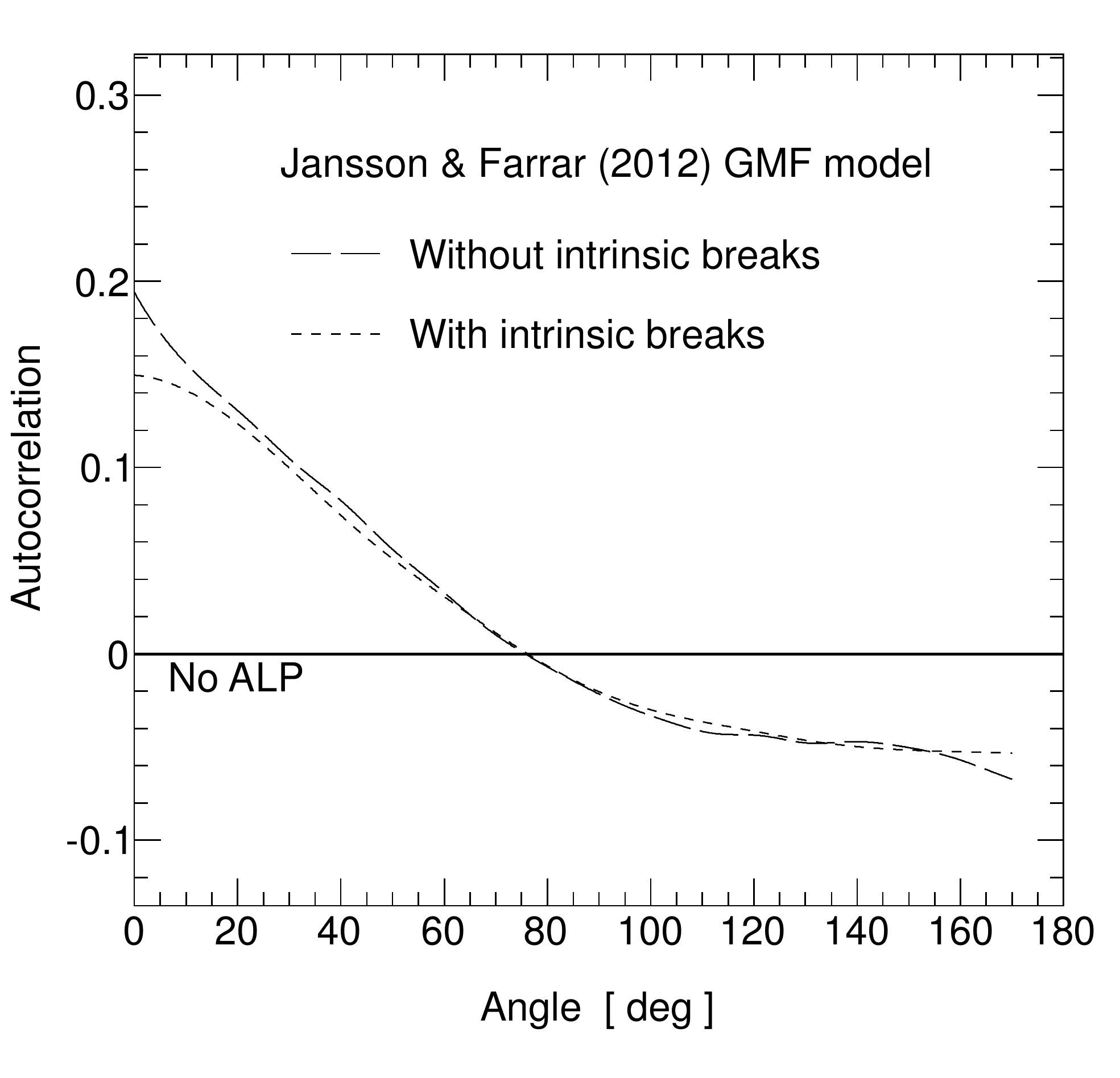}
\caption{Autocorrelation ALP predictions assuming $g_{\gamma a} = 5\times10^{-11}$ GeV$^{-1}$ and the GMF of~\cite{2012ApJ...757...14J} with and without intrinsic spectral breaks (see text for details).}
\label{fig:model_break}
\end{figure}

It could be that the presence of intrinsic curvature of the source spectra, that would be seen as intrinsic cut offs due to the finite energy resolution, change the results. Such an effect would add to the difference in spectral indices, in a way unrelated to the transparency scenario and could eventually blur the observable. To study the impact of the intrinsic curvature of spectra on the test, the probability density function of intrinsic spectral break between the HE and VHE domains, given by~\cite{2013A&A...554A..75S} is used. This probability density function is computed by simulating blazar spectral energy distributions (SEDs) following a synchrotron self Compton model with one zone. This model successfully reproduces the SED of BL-Lac objects, which will form the bulk of the extragalactic sources detected by CTA. A break between the HE and VHE spectral indices of each source is randomly added, following this probability density function. Figure~\ref{fig:model_break} shows the autocorrelation prediction computed when spectral breaks are randomly added. The results are compared to the prediction when no break is considered, for $g_{\gamma a} = 5\times10^{-11}$ GeV$^{-1}$ and the GMF of~\cite{2012ApJ...757...14J}. The effect is still sizable in spite of the addition of intrinsic spectral break, which is therefore not a limitation to the test. As for the knowledge of the redshifts, the fact that the autocorrelation method still works here is understandable because intrinsic spectral breaks change the distribution of the $\Delta\Gamma$ in just the same manner anywhere on the sky, without strongly modifying the correlations.

\begin{figure}[h]
\centering
\includegraphics[width=0.5\textwidth]{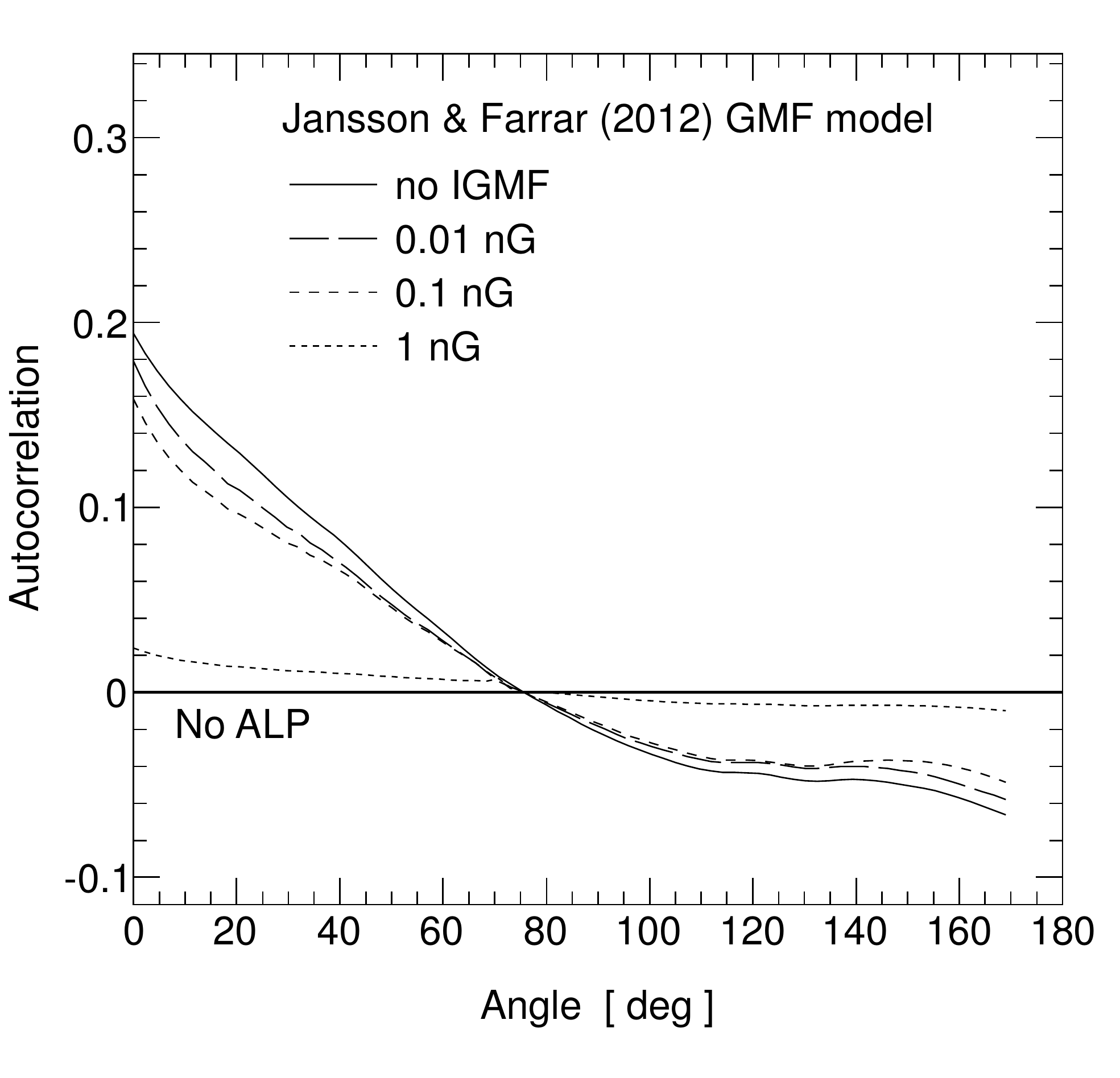}
\caption{Autocorrelation ALP predictions for different values of the IGMF strength, assuming $g_{\gamma a} = 5\times10^{-11}$ GeV$^{-1}$ and the GMF of~\cite{2012ApJ...757...14J}.\label{fig:model_igmf}}
\end{figure}

In Sec.~\ref{sec:6}, it is assumed that the strength of the IGMF is low enough so that it could not efficiently mix photons and ALPs. However, in the case of a strong IGMF, close to the upper bound of 1 nG for large turbulence scales of order of 1 Mpc, the mixing is strong enough to influence the test. 
To study the effect of a strong IGMF, the autocorrelation model is computed for different values of the IGMF strength, assuming a coherence length of 1 Mpc. The result, for $g_{\gamma a} = 5\times10^{-11}$ GeV$^{-1}$ and the GMF of~\cite{2012ApJ...757...14J} is shown in Fig.~\ref{fig:model_igmf}. For a large IGMF strength of 1 nG, the autocorrelation pattern is significantly reduced. If the look-elsewhere effect induced by the finite sample of 500 sources expected for CTA is included (see Fig.~\ref{fig:corr_simu}), the autocorrelation is compatible with zero. In that case, the test cannot conclude. For lower values of the IGMF, the sensitivity is retrieved.

In Fig.~\ref{fig:sensitivity}, the region of the parameter space that could explain the anomalous spectra claimed in~\cite{Horns:2012fx} is shown (from~\cite{2013PhRvD..87c5027M}). The test proposed in this study covers a significant part of this parameter space, another part being already excluded by constraints from spectral irregularities at TeV energies~\cite{HESS_ALP}. Independent studies are required to cover all of this parameter space region. Note that the expected CTA sample of 500 sources used in this study will only be fully complete at the end of operation of CTA. Prospected instruments such as IAXO~\cite{Irastorza:2012qf} will also be able to probe this parameter space and perhaps then seal the issue before the end of the full life time of CTA. Nevertheless, the study on the blazar population expected to be detected by CTA~\cite{2013APh....43..215S} shows that in only two years of operation, CTA should detect at least 140 extragalactic sources with known redshifts, thus already more than the triple of the current sample, enabling the use of the test. The CTA simulations used in the previous sections are performed with the array B configuration. In that case the effective area is larger at low energy, and this provides the best prospect in terms of absolute number of detected sources. To estimate the importance of the array configuration, the analysis has been redone completely using the array I. In that case, compared to the B configuration, the effective area is larger at higher energies, at the expense of a larger threshold. As a consequence, less AGN (370) are expected to be detected in that configuration (see Fig.~7 of~\cite{2013APh....43..215S}). For the sensitivity to the ALP transparency signal through the autocorrelation, the lower number of sources is counterbalanced by a better determination of $\Delta\Gamma$, in particular thanks to the larger number of collected photons at high energy. Eventually these two antagonist effects compensate, as the sensitivity on the coupling is found to be $3.0\times 10^{-11}\;\rm GeV^{-1}$ in the case of the array I, compared to $2.9\times 10^{-11}\;\rm GeV^{-1}$ in the case of the array B. Finally, note that when CTA will be running, archival {\it Fermi}-LAT data will be available and could be used to improve the analysis, in particular by further lowering the energy threshold and improve the $\Gamma_HE$ measurements.

\section{Conclusions}

A new observable related to the ALP-induced anomalous transparency of the universe is proposed. It is based on the measurement of the differences of spectral indices between the HE and the VHE ($\Delta\Gamma$). If the scenario of transparency induced by ALPs is indeed at work, the effect is dependent of the integrated GMF on the line of sight, which offers a possibility to test it through the $\Delta\Gamma$ autocorrelation. The current sample of extragalactic sources detected at very high energies is not large enough for this test to be conclusive. It is shown that CTA, with its expected large sample of 500 extragalactic sources, will be sensitive to the autocorrelation observable. The test is robust with respect to the knowledge of the redshifts, intrinsic spectral breaks and requires that the IGMF is weaker than 1 nG. The detection by CTA of an anomaly autocorrelation signal would be a strong indication for the existence of ALPs. A final test would be to cross-correlate the anomaly pattern with specific GMF models. The proposed observable is complementary to the search for irregularities in the source spectra, and the search for an opacity anomaly based on the magnitude and statistics of the spectral hardening.

\acknowledgments

We are very grateful to Pasquale Serpico, who first proposed a similar idea. We would like to thank Konrad Bernlöhr for his help with the CTA instrument response functions, and the HESS team in Saclay within which this analysis has be thoroughly discussed. Part of this work was supported by the French national program PNHE and the ANR project CosmoTeV.

\bibliographystyle{JHEP}
\bibliography{wouters_brun.bib}

\end{document}